\documentclass[12pt]{amsart}
\usepackage[margin=1in]{geometry}
\usepackage{amsfonts}
\usepackage{amsmath}
\usepackage{amssymb}
\usepackage{amsthm}
\usepackage{graphicx}
\usepackage{amsfonts}
\usepackage{color}

\newtheorem{theorem}{Theorem}[section]
\newtheorem{lemma}[theorem]{Lemma}
\newtheorem{corollary}[theorem]{Corollary}
\newtheorem{proposition}[theorem]{Proposition}
\newtheorem{remark}[theorem]{Remark}

\newtheorem{definition}[theorem]{Definition}

\newcommand{\cO}{{\mathcal O}}

\newcommand{\R}{{\mathord{\mathbb R}}}
\newcommand{\C}{{\mathord{\mathbb C}}}
\newcommand{\Z}{{\mathord{\mathbb Z}}}

\newcommand{\E}{{\mathord{\mathbb E}}}
\newcommand{\PP}{{\mathord{\mathbb P}}}

\newcommand{\be}{\begin{equation}}
\newcommand{\ee}{\end{equation}}
\newcommand{\bea}{\begin{eqnarray}}
\newcommand{\eea}{\end{eqnarray}}

\numberwithin{equation}{section}

\begin{document}

\title[Oscillator Systems]{Quantum Harmonic Oscillator Systems with Disorder}

\author[B. Nachtergaele]{Bruno Nachtergaele$^1$}
\thanks{B.\ N.\ was supported in part by NSF grant DMS-1009502}
\address{$^1$ Department of Mathematics\\ University of California, Davis\\ Davis, CA 95616, USA}
\email{bxn@math.ucdavis.edu}

\author[R. Sims]{Robert Sims$^2$}
\thanks{R.\ S.\ was supported in part by NSF grants DMS-0757424 and DMS-1101345}
\address{$^2$ Department of Mathematics\\
University of Arizona\\
Tucson, AZ 85721, USA}
\email{rsims@math.arizona.edu}

\author[G. Stolz]{G\"unter Stolz$^3$}
\thanks{G.\ S.\ was supported in part by NSF grant DMS-1069320.}
\address{$^3$ Department of Mathematics\\
University of Alabama at Birmingham\\
Birmingham, AL 35294 USA}
\email{stolz@math.uab.edu}

\date{}

%
\begin{abstract}
We study many-body properties of quantum harmonic oscillator lattices with disorder.
A sufficient condition for dynamical localization, expressed as a zero-velocity Lieb-Robinson
bound, is formulated in terms of the decay of the eigenfunction correlators for an effective
one-particle Hamiltonian. We show how state-of-the-art techniques for proving Anderson
localization can be used to prove that these properties hold in a number of standard models.
We also derive bounds on the static and dynamic correlation functions at both zero
and positive temperature in terms of one-particle eigenfunction correlators. In particular, we
show that static correlations decay exponentially fast if the corresponding effective
one-particle Hamiltonian exhibits localization at low energies, regardless of whether there is
a gap in the spectrum above the ground state or not. Our results apply to finite as well as to
infinite oscillator systems. The eigenfunction correlators that appear are more general than
those previously studied in the literature. In particular, we must allow for functions of the
Hamiltonian that have a singularity at the bottom of the spectrum. We prove exponential bounds
for such correlators for some of the standard models.
\end{abstract}

\maketitle

%
%
%

\section{Introduction}

Oscillator lattice systems are the standard model for the vibrational
degrees of freedom, known as phonons, in crystal lattices. These phonons
interact with the other degrees of freedom, such as spins and
electrons, in ways that often significantly modify their behavior. The two-point
correlation functions in the ground state and thermal states of the lattice,
e.g. $\langle q_x q_y\rangle$ and $\langle p_x p_y\rangle$, and their
time-dependent analogues, are key quantities to understand the role of the
lattice variables in the structure and transport properties of materials.

For a pure regular crystal in the harmonic approximation, finding the correlation functions
is a standard exercise. It has also long been understood, however,
that to describe a basic phenomenon such as heat conduction, it is important to
go beyond the pure harmonic lattice and to consider anharmonicities \cite{Lippi2000}
and randomness \cite{Rieder1967}.
Anharmonicities lead to non-linear phonon interactions, as studied,
e.g., in \cite{Spohn}. In this work we are interested in the effects of randomness,
which could e.g.\ occur in a crystalline material layered on a substrate.

The main effect of randomness is that it may lead to localization of the phonons
at certain energies, which means that a portion of the energy spectrum consists
of eigenvectors that decay exponentially fast at large distances. When localization
occurs at all energies the physical effect is {\em dynamical localization}, i.e.,
the absence of propagation of disturbances in the system. This phenomenon has been
studied extensively in one-particle models such as the Anderson model, see
\cite{Stollmann, Kirsch, Klein, Stolz11} for surveys and more references.
The mathematical study of localization in systems with many degrees of freedom has
only recently begun, e.g.\ \cite{ChulaevskySuhov1, ChulaevskySuhov2, AizenmanWarzel, HSS11}.

There is no consensus on how to mathematically quantify dynamical localization in the many-body context.
With this as motivation, we will presently, as indicated above, consider random oscillator systems.
For such systems, much can be analyzed directly and in fact, we will prove two main
estimates which, under certain assumptions, suggest that the system is dynamically localized.
First, we give conditions under which the models satisfy a zero velocity Lieb-Robinson bound.
Next, we prove that time-dependent correlations (both ground state and thermal) decay exponentially in space
with no gap assumption. Below, we briefly describe both of these results.

Our first results demonstrate a condition under which the random oscillator systems we
consider satisfy a strong form of dynamical localization; namely, a zero velocity Lieb-Robinson bound
(in disorder average). Generally, Lieb-Robinson bounds are deterministic and show that the
Heisenberg evolution (corresponding to a
Hamiltonian comprised of sufficiently short range interactions) of a local observable
remains essentially local. More precisely, they prove that given an observable initially supported
in a finite set $X \subset \mathbb{Z}^d$, the dynamic evolution of this observable to time $t$ has,
up to exponentially small corrections, support contained in a ball centered at $X$ of radius $v|t|$. The number
$v>0$ is called the group velocity or velocity of propagation for the system.
Results of this type were originally proven by Lieb and Robinson \cite{LR1} in the context of quantum spin systems,
and they have recently been generalized \cite{NS1,H+K}, in particular to the setting of oscillators systems,
see e.g.\ \cite{NRSS, CSE,Nourr}.

Our results here, see e.g.\ Theorems~\ref{thm:sdlweylcom} and \ref{thm:sdlp+qcom} for finite oscillator systems and Theorem~\ref{thm:infvoldynloc} for an extension to infinite systems,
give assumptions under which random oscillator systems satisfy a Lieb-Robinson bound, on average,
uniformly in time, i.e. a bound with $0$ velocity of propagation. Similar results were recently obtained in
\cite{HSS11} for the random XY chain.

The condition for the above results to hold constitutes our main new technical input.
It is described in terms of estimates on certain {\it singular eigenfunction correlators} corresponding
to the effective one-particle Hamiltonian. In the single particle theory of random systems, eigenfunction
correlators are a well-known tool for deriving results on dynamical localization, see e.g.\
\cite{Aizenman94, ASFH, Stolz11}. Here the new insight is that,
for the effective models corresponding to random oscillator systems, {\it singular} eigenfunction correlators
need to be estimated, and we demonstrate, see specifically Appendices~\ref{app:Anderson} and \ref{app:AndersonIV}, that the desired
bounds can indeed be proven in certain well-studied systems.

Our next results concern correlation decay. It is well-known that, deterministically, gapped ground states
satisfy exponential clustering, i.e.\ the correlations of observables in gapped ground states decay
exponentially in the distance between the supports of the observables, see \cite{NS1,H+K} (spin systems)
and \cite{SCW, CramerEisert, NRSS}  (oscillator systems) for proofs of this fact.

For random oscillator systems, we give conditions under which ground state correlations
satisfy exponential clustering, on average; even in cases where the gap closes. Again, the conditions for
these results also involve estimating certain singular eigenfunction correlators. We note that such a situation,
one for which exponential clustering holds without the existence of a uniform gap,
can be referred to as a regime with a {\it mobility gap}, and this has been discussed in the literature,
see e.g. \cite{Hastings}. Specifically, our results show that, under
certain assumptions, exponential clustering holds, on average, for both static and dynamic
correlations in ground states and also in thermal states.

We now give a detailed outline of the contents of this paper.
In Section~\ref{Sec:OscSystems}, we introduce a class of oscillator systems depending explicitly on
three sets of parameters: the masses, the spring constants, and the interaction strengths.
To establish our notation, we briefly review the well-known procedure by which these models can be diagonalized
and comment on some important consequences. Here, as well as in the following Sections~\ref{sec:dlc} to \ref{sec:applications}, we consider {\it finite} (but arbitrarily large) oscillator systems.

In Section \ref{sec:dlc}, we prove our strongest form of localization for the dynamics corresponding to
random oscillator systems. This localization is expressed in terms of a zero-velocity Lieb-Robinson bound, in average,
and it follows from an appropriate exponential decay of the eigenfunction correlators corresponding to the
effective one-particle Hamiltonian. We show that these results apply to two classes of observables:
in Section~\ref{subsec:comweyl} we consider Weyl operators, while in Section~\ref{subsec:comp+q}
our bounds for the positions and momenta are discussed.

In Sections~\ref{sec:gsc} and \ref{sec:tsc}, we present our results on dynamic and static correlation decay,
again for both Weyl operators as well as the positions and momenta. Section~\ref{sec:gsc} discusses
correlations in the ground state, whereas Section~\ref{sec:tsc} demonstrates our results for thermal states.

In Section~\ref{sec:applications}, we focus on our main application --- the case in which the effective
one-particle Hamiltonian becomes the Anderson model. Here we illustrate how the bounds on the
many-body system which we derived in Sections~\ref{sec:dlc}, \ref{sec:gsc}, and \ref{sec:tsc} depend on
certain well-studied localization regimes for the Anderson model, e.g.
large disorder and one-dimension. We also describe what our results show in cases where
dynamical localization has only been proven at the bottom of the spectrum.

The extension of our results from finite to infinite oscillator systems is the content of Section~
\ref{sec:infinitevolume}. We focus on the case where the effective one-particle Hamiltonian is
the infinite volume Anderson model.  Again we obtain zero-velocity Lieb-Robinson bounds as
well as exponential decay of ground state and thermal state correlations.

In Section~\ref{sec:conclusion} we provide a brief non-technical summary of our new results.

Two appendices contain results on Anderson localization for the effective one-particle
operators associated with the oscillator system Hamiltonians, separate for finite volume
(Appendix~\ref{app:Anderson}) and infinite volume (Appendix~\ref{app:AndersonIV}). Many of
these results, in particular those on singular eigenfunction correlators, have not previously
appeared in the literature. Thus we will present them with detailed proofs. We think that these
results are of independent interest in the theory of Anderson localization.

A number of the conditional results in this paper, see specifically those in Sections~\ref{sec:dlc}, \ref{sec:gsc},
and \ref{sec:tsc}, apply to much more general systems (oscillator systems over more general graphs, more
general interactions, see Remark~\ref{rem:2} below). However, for the sake of clarity, we have compromised
obvious extensions to illustrate our main techniques. Moreover, our primary application to the case of Anderson-type single particle Hamiltonians is of the form discussed, and more general results on dynamical
localization for the corresponding single particle models are generally not known.

A related topic of great current interest are the so-called Area Law bounds
for the entropy of entanglement in the ground state of multi-dimensional
systems. The techniques of this paper can be used to obtain such a bound
for the random oscillator lattices considered here, under the assumption
that there is a mobility gap (but not necessarily a gap) above
the ground state. We will report on that result in a separate article \cite{NSS}.

\section*{Acknowledgements}

This collaboration was made possible by a Summer 2012 {\em Research in Teams} project at
the Erwin Schr\"o\-dinger International Institute for Mathematical Physics, University of Vienna,
support from which we gratefully acknowledge. We also acknowledge useful discussions with
Michael Aizenman, Sven Bachmann and Jeff Schenker on several important aspects of this work.
G.\ S.\ is grateful for hospitality and financial support at the University of California, Davis and the
University of Arizona, where this work was initiated during a sabbatical leave in Fall 2011.

%
%
%

\section{Systems of coupled harmonic quantum oscillators} \label{Sec:OscSystems}

We will consider oscillator models defined over $\mathbb{Z}^d$, $d \geq 1$ an integer. These we describe
in terms of three sequences of parameters: the masses $\{ m_x \}$,
the spring constants $\{ k_x \}$, and the couplings $\{ \lambda_{x,y} \}$.
Throughout this work, we will assume the following.

\vspace{.2cm}

{\bf Assumption A.1:} For each $x \in \mathbb{Z}^d$, $m_x>0$, $k_x \geq 0$, and
for each two element set $\{ x, y \} \subset \mathbb{Z}^d$ of nearest neighbor sites
(e.g.\ $|x-y|=1$, where $|\cdot|$ will denote the $1$-norm), $\lambda_{x,y} \geq 0$.
Moreover, we assume there are numbers $m_{\rm min}$, $m_{\rm max}$, $k_{\rm max}$, and $ \lambda_{\rm max}$
such that $0< m_{\rm min} \leq m_x \leq m_{\rm max}$, $0 \leq k_x \leq k_{\rm max}$, and $0 \leq \lambda_{x,y} \leq \lambda_{\rm max}$.

\vspace{.2cm}

In terms of these sequences, finite-volume harmonic Hamiltonians are defined as follows.
Fix an integer $L \geq 1$ and a cubic set $\Lambda_L = [-L,L]^d \cap \Z^d$.
Set
\begin{equation} \label{generalHam}
H_L = \sum_{x\in \Lambda_L} \left( \frac{1}{2m_x} p_x^2 + \frac{k_x}{2} q_x^2 \right) + \sum_{\{x,y\} \subset \Lambda_L
\atop  |x-y|=1} \lambda_{x,y} (q_x-q_y)^2
\end{equation}
to be the harmonic Hamiltonian on $\Lambda_L$ which acts on the Hilbert space
\begin{equation}
{\mathcal H}_L = L^2(\R^{\Lambda_L},dq) = \bigotimes_{x\in \Lambda_L} L^2(\R, dq_x),
\end{equation}
with $q=(q_x)_{x\in \Lambda_L}$ denoting the space variables.
Note here that we use subsets $\{x,y\} \subset \Lambda_L$ rather than pairs $(x,y) \in \Lambda_L \times \Lambda_L$, meaning that $\{\{x,y\}: |x-y|=1\}$ labels the {\it undirected} edges of the next-neighbor graph associated with $\Lambda_L$. Moreover, we also use $q_x$ to denote the position
operators, meaning the multiplication operators by $q_x$, and by $p_x =-i\partial/\partial q_x$ we denote the momentum operators.
In this case, $H_L$ describes a $d$-dimensional system of one-dimensional harmonic oscillators that are coupled by nearest neighbor quadratic
interactions. We consider finite systems of side-length $2L+1$, but will be interested in results (and bounds) which hold uniformly in the size of the system.
This is one reason why we assume the uniform bounds on the parameters found in Assumption {\bf A.1}.

By standard results, e.g.\ \cite{ReedSimon2}, each $q_x$, $p_x$, and also $H_L$ is essentially self-adjoint on $C_0^{\infty}(\R^{\Lambda_L})$, the smooth functions of
compact support, as well as on the Schwarz space of rapidly decreasing functions over $\R^{\Lambda_L}$. From now on we will use $H_L$ to denote the self-adjoint closure of this operator. Clearly, $H_L$ is non-negative.

As is well-known, it is possible to describe the eigenvalues and eigenfunctions of $H_L$ quite explicitly.
This is due to the fact that the harmonic Hamiltonian $H_L$ can be reduced to a system of uncoupled
harmonic oscillators (or free Boson system) via a quantum canonical (or Bogoliubov) transformation, see, e.g., \cite{DeGosson} or \cite{BratRob} for background. We review this procedure in order to introduce some relevant notation. To start, we express $H_L$ in the form
\begin{equation} \label{matrixrep}
H_L = ( q^T, p^T)\; {\mathbb H}_L \left( \begin{array}{c} q \\ p \end{array} \right).
\end{equation}
where
\begin{equation} \label{matrixform}
{\mathbb H}_L = \left( \begin{array}{cc} h_{0,L} & 0 \\ 0 & \mu \end{array} \right).
\end{equation}
Here we view $q= (q_x)$ and $p= (p_x)$ as column vectors indexed by $x\in \Lambda_L$, with the transposes $q^T$ and $p^T$ denoting the corresponding row vectors. With this convention (\ref{matrixrep}) can be understood in the sense of standard matrix multiplication. By $\mu$ we denote the diagonal matrix $\mu_{x,y} = \frac{1}{2m_x} \delta_{x,y}$, which we will generally think of as a multiplication operator on $\ell^2(\Lambda_L)$ (omitting the $L$-dependence in our notation). Thus we may write $\mu_{x,y} = \langle \delta_x, \mu \delta_y \rangle$, with $\{\delta_x\}_{x\in \Lambda_L}$ the canonical basis in $\ell^2(\Lambda_L)$.

The matrix-elements of the operator $h_{0,L}$ in $\ell^2(\Lambda_L)$ are given by
\begin{equation} \label{eq:matrixelemh0}
\langle \delta_x, h_{0,L} \delta_y \rangle = \left\{ \begin{array}{ll} \frac{k_x}{2} + \sum_{u:|u-x|=1} \lambda_{x,u}, & \mbox{if $x=y$}, \\ - \lambda_{x,y}, & \mbox{if $|x-y|=1$}, \\ 0, & \mbox{else.} \end{array} \right.
\end{equation}
The associated form is given by
\begin{equation} \label{eq:h0form}
\langle f, h_{0,L} \, g \rangle =   \sum_{\{x,y\} \subset \Lambda_L \atop |x-y|=1} \lambda_{x,y} \overline{(f(y)-f(x))} (g(y)-g(x))  + \sum_{x\in \Lambda_L} \frac{k_x}{2} \overline{f(x)} g(x)
\end{equation}
for all $f, g \in \ell^2(\Lambda_L)$. This shows that $h_{0,L}$ is non-negative. It also follows from (\ref{eq:matrixelemh0}) and {\bf A.1} that, uniformly in $L$,
\begin{equation} \label{eq:normboundh0}
\|h_{0,L}\| \le 4d \lambda_{\rm max} + \frac{1}{2} k_{\rm max} \,
\end{equation}

Our second assumption is as follows.

\vspace{.2cm}

{\bf Assumption A.2:} We assume that $h_{0,L}$ is positive definite.

\vspace{.2cm}

Since $h_{0,L} \ge \frac{1}{2} \min_{x\in \Lambda_L} k_x$, {\bf A.2} holds for all $L$ if there is a number $k_{\rm min}$ with $0 < k_{\rm min} \leq k_x$.
There are also other conditions which guarantee that {\bf A.2} holds.  Assumption {\bf A.2} is sufficient to guarantee that
$H_L$ has purely discrete spectrum and a non-degenerate, normalized ground state $\Omega_L$. This will be clear as a
result of the diagonalization, which we continue with below, see e.g.\ Remark~\ref{rem:1}.

The position and momentum operators $q_x$ and $p_x$, on suitable domains, satisfy the self-adjoint form of the canonical commutation relations (CCR)
\begin{equation} \label{CCRsa}
[p_x,p_y] = [q_x,q_y] = 0 \quad \mbox{and} \quad [q_x,p_y] = i\delta_{x,y} I.
\end{equation}
A canonical transformation is a linear change of position and momentum operators implemented via a symplectic matrix, see \cite{DeGosson}. More precisely, define $Q=(Q_k)$, $P=(P_k)$, $k=1,\ldots,|\Lambda_L|$, by
\begin{equation} \label{modifiedqp}
\left( \begin{array}{c} Q \\ P \end{array} \right) = S^{-1} \left( \begin{array}{c} q \\ p \end{array} \right),
\end{equation}
where $S$ is a $2|\Lambda_L|$-by-$2|\Lambda_L|$ symplectic matrix, i.e.\
\begin{equation} \label{defsymplectic}
SJS^T = J, \quad J:= \left( \begin{array}{cc} 0 & -I \\ I & 0 \end{array} \right).
\end{equation}
One can check that the modified position and momentum operators still satisfy
\begin{equation} \label{CCRQP}
[P_j,P_k] = [Q_j,Q_k]=0 \quad \mbox{and} \quad [Q_j, P_k] = i\delta_{j,k} I,
\end{equation}
and can be used to express the quadratic Hamiltonian in the form
\begin{equation} \label{newmatrixrep}
H_L = ( Q^T, P^T)\; S^T {\mathbb H}_L S \left( \begin{array}{c} Q \\ P \end{array} \right).
\end{equation}

Every quadratic Hamiltonian of the form (\ref{matrixrep}) with a real symmetric matrix ${\mathbb H}_L$ can be diagonalized in the sense that a symplectic matrix $S$ exists such that $S^T {\mathbb H}_L S$ in (\ref{newmatrixrep}) becomes diagonal. This is the content of Williamson's theorem, e.g.\ Section 8.3 in \cite{DeGosson}. In our case (\ref{matrixform}) this can be done explicitly in two simple steps.

For the first step let
\begin{equation}
S_1 = \left( \begin{array}{cc} \mu^{1/2} & 0 \\ 0 & \mu^{-1/2} \end{array} \right).
\end{equation}
One checks that $S_1$ is symplectic  and that
\begin{equation}
S_1^T {\mathbb H}_L S_1 = \left( \begin{array}{cc} h_L & 0 \\ 0 & I \end{array} \right),
\end{equation}
where
\begin{equation} \label{onepartham}
h_L = \mu^{1/2} h_{0,L} \, \mu^{1/2}.
\end{equation}

By our assumptions, the real symmetric matrix $h_L$ is positive definite and satisfies the norm bound
\begin{equation} \label{eq:normboundh}
\|h_L\| \le \|\mu\| \|h_{0,L}\| \le \frac{1}{2m_{\rm min}} (4d \lambda_{max} + \frac{1}{2} k_{\rm max}).
\end{equation}

In the second step $h_L$ is diagonalized in terms of a real orthogonal matrix $\cO$, i.e.,
\begin{equation} \label{diagtildeh}
\cO^T \,h_L \,\cO =  \gamma^2 \, ,
\end{equation}
with $ \gamma^2 = \mbox{diag}( \gamma_k^2)$ (and, for later reference, $\gamma = \mbox{diag}(\gamma_k)$). We have set $\gamma_k^2$ to be the eigenvalues of
 $h_L$ counted with multiplicity, taking $\gamma_k >0$ for each $k =1, 2, \ldots, | \Lambda_L|$.

Let
\begin{equation}
S_2 := \left( \begin{array}{cc} \cO & 0 \\ 0 & \cO \end{array} \right),
\end{equation}
which is symplectic. Thus
\begin{equation} \label{fulltransform}
S=S_1 S_2 = \left( \begin{array}{cc} \mu^{1/2} \,\cO & 0 \\ 0 & \mu^{-1/2} \,\cO \end{array} \right)
\end{equation}
is symplectic and
\begin{equation}
S^T \left( \begin{array}{cc} h_{0,L} & 0 \\ 0 & \mu \end{array} \right) S = \left( \begin{array}{cc} \gamma^2 & 0 \\ 0 & I \end{array} \right),
\end{equation}
which we had set out to accomplish.

In terms of $Q$ and $P$ from (\ref{modifiedqp}) the harmonic Hamiltonian can be rewritten as
\begin{eqnarray} \label{freeoscillators}
H_L & = & (Q^T, P^T) \left( \begin{array}{cc} \gamma^2 & 0 \\ 0 & I \end{array} \right) \left( \begin{array}{c} Q \\ P \end{array} \right) \\ \label{freeoscillators2}
& = & \sum_k (P_k^2 + \gamma_k^2 Q_k^2).
\end{eqnarray}
This has the form of a system of uncoupled harmonic oscillators (or free Boson system) which can be explicitly diagonalized: Let
\begin{equation} \label{eq:bkeq}
B_k = \sqrt{\frac{\gamma_k}{2}} Q_k + \frac{i}{\sqrt{2\gamma_k}} P_k, \quad k = 1,\ldots, | \Lambda_L|,
\end{equation}
a set of operators which together with their adjoints $B_k^*$ satisfy CCR in the creation-annihilation-operator form
\begin{equation} \label{CCRB}
[B_j, B_k] = [B_j^*, B_k^*] = 0, \quad [B_j, B_k^*] = \delta_{j.k} I.
\end{equation}
One gets
\begin{equation} \label{diagH}
H_L = \sum_k \gamma_k (2B_k^* B_k + I).
\end{equation}
It follows that
\begin{itemize}
\item $H_L$ has a unique normalized ground state $\Omega_L$, with ground state energy $\gamma_0 = \sum_k \gamma_k$, which is characterized by $B_k \Omega_L = 0$ for all $k$.
\item An orthonormal basis of eigenvectors for $H_L$ is given by
\begin{equation} \label{eq:evecs}
\prod_{j=1}^{|\Lambda_L|}  \frac{1}{\sqrt{\alpha_j!}} (B_j^*)^{\alpha_j}  \Omega_L,
\end{equation}
where $\alpha_j$, $j\in \{1,\ldots, |\Lambda_L|\}$, are arbitrary non-negative integers. The corresponding eigenvalues are
\begin{equation} \label{eq:evalsH}
\gamma_0 + 2 \sum_{j=1}^{|\Lambda_L|} \alpha_j \gamma_j.
\end{equation}
In particular, the ground state gap is $2\min_k \gamma_k$.
\end{itemize}

For many calculations, it is convenient to express the original position and momentum operators in terms of these
operators $B_k$ which diagonalize $H_L$. Note that, with the explicit symplectic transformation in (\ref{fulltransform}), (\ref{modifiedqp}) can
be rewritten as
\begin{equation}
\left( \begin{array}{c} q \\ p \end{array} \right) = \left( \begin{array}{cc} \mu^{1/2} \mathcal{O} & 0 \\ 0 & \mu^{-1/2} \mathcal{O} \end{array} \right) \left( \begin{array}{c} Q \\ P \end{array} \right) \, .
\end{equation}
Recalling (\ref{eq:bkeq}), we see that the $B$'s are defined as
\begin{equation}
\left( \begin{array}{c} B \\ B^* \end{array} \right) = \frac{1}{\sqrt{2}} \left( \begin{array}{cc} \gamma^{1/2} & i \gamma^{-1/2} \\ \gamma^{1/2} &-i \gamma^{-1/2} \end{array} \right) \left( \begin{array}{c} Q \\ P \end{array} \right) \, ,
\end{equation}
and therefore,
\begin{equation} \label{eq:q+pvecB}
\left( \begin{array}{c} q \\ p \end{array} \right) = \frac{1}{ \sqrt{2}} \left( \begin{array}{cc} \mu^{1/2} \mathcal{O} \gamma^{-1/2} & 0 \\ 0 & \mu^{-1/2} \mathcal{O} \gamma^{1/2} \end{array} \right)
\left( \begin{array}{cc}  1 &  1 \\ -i & i \end{array} \right) \left( \begin{array}{c} B \\ B^* \end{array} \right) \, .
\end{equation}
An explicit expression for the evolution of (\ref{eq:q+pvecB}) under the dynamics generated by $H_L$ will also be important. We briefly discuss this next.

By Stone's Theorem,
given an initial state $\psi \in D(H_L)$, the unique solution of the Schr\"odinger equation
\begin{equation}
i \partial_t \psi(t) = H_L \psi(t) \quad \mbox{with} \quad \psi(0) = \psi \, ,
\end{equation}
is given by $\psi(t) = e^{-itH_L} \psi$, where the unitary operators $e^{-itH_L}$ are defined via the functional calculus for self-adjoint operators. An alternative description of this dynamics is
in terms of the so-called Heisenberg picture. Here, for example, one can define a one-parameter
group of automorphisms $\tau_t$ on $\mathcal{B}( \mathcal{H}_L)$ by setting
\begin{equation} \label{eq:harmdyn}
\tau_t(A) = e^{it H_L}Ae^{-itH_L} \quad \mbox{for all} \quad A \in \mathcal{B}( \mathcal{H}_L) \, .
\end{equation}
$\tau_t$ is called the Heisenberg dynamics generated by $H_L$; it depends on $L$, but we suppress
that in our notation.

One can extend this notion of dynamics to certain unbounded operators on $\mathcal{H}_L$.
In fact, appealing again to Stone's Theorem, one can check that the function $b_k(t) = \tau_t(B_k)$
is the unique solution of $b_k'(t) = -2i \gamma_k b_k(t)$ with $b(0) = B_k$. (The previous equation holds, in the
strong sense, on $D(H_L)$ and the solution can be extended by linearity to the domain
of $B_k$.) In this case, we conclude
\begin{equation} \label{eq:bkevolv}
\tau_t(B_k) = e^{-2i \gamma_k t} B_k \quad \mbox{and similarly} \quad \tau_t(B_k^*) = e^{2i \gamma_k t} B_k^* \, .
\end{equation}
Applying (\ref{eq:bkevolv}) to (\ref{eq:q+pvecB}) we find that
\begin{equation} \label{eq:matp+qtev}
\left( \begin{array}{c} \tau_t(q) \\ \tau_t(p) \end{array} \right) = \frac{1}{ \sqrt{2}} \left( \begin{array}{cc} \mu^{1/2} \mathcal{O} \gamma^{-1/2} & 0 \\ 0 & \mu^{-1/2} \mathcal{O} \gamma^{1/2}\end{array} \right)
\left( \begin{array}{cc}  e^{-2i \gamma t} &  e^{2i \gamma t} \\ -i e^{-2i \gamma t} & i e^{2i \gamma t} \end{array} \right) \left( \begin{array}{c} B \\ B^* \end{array} \right) \, .
\end{equation}

We end this section with a few remarks.

\begin{remark} \label{rem:1} While it is convenient to express the above reduction of $H_L$ to a free Boson system in terms of the canonical transformation formalism, one may equivalently proceed by expressing the above transformations as unitary equivalences: Consider unitary operators $U_1$ and $U_2$ on $L^2(\R^{\Lambda_L})$ defined by $(U_1 f)(q) = cf(\mu^{1/2}q)$ and $(U_2 f)(q) = f(\cO q)$. Here $\mu$ and $\cO$ are the diagonal and orthogonal matrices defined above and $c=\prod_x (2m_x)^{1/4}$ normalizes $U_1$ to be unitary. Let $ U := U_2 U_1$. $U_1$ rescales all masses to be $1/2$ and $U_2$ diagonalizes the quadratic potential, so that what we have shown above amounts to
\begin{equation}
U H_L U^* = -\Delta + \sum_k \gamma_k^2 q_k^2,
\end{equation}
meaning that $H_L$ is unitarily equivalent to a system of uncoupled harmonic oscillators. This has eigenvalues given by (\ref{eq:evalsH}) with eigenvectors (\ref{eq:evecs}) taking the explicit form of Hermite functions, thus also showing completeness. From this point of view one gets the modified positions and momenta satisfying (\ref{freeoscillators2}) as
\begin{equation}
Q_k = U^* q_k U, \quad P_k = U^* p_k U.
\end{equation}

\end{remark}

\begin{remark} \label{rem:2}
The above considerations can easily be modified to apply to more general quadratic Hamiltonians (including multi-dimensional oscillators) of the form
\begin{equation}
H = ( q^T, p^T)\;  \left( \begin{array}{cc} h_1 & 0 \\ 0 & h_2 \end{array} \right) \left( \begin{array}{c} q \\ p \end{array} \right),
\end{equation}
with positive matrices $h_1$ and $h_2$. Such cases are considered in \cite{CramerEisert}. The symplectic matrix in (\ref{fulltransform}) is generalized by

\begin{equation}
S= \left( \begin{array}{cc} h_2^{1/2} \,\cO & 0 \\ 0 & h_2^{-1/2} \,\cO \end{array} \right),
\end{equation}
where $\cO$ is the diagonalizing orthogonal matrix for $h := h_2^{1/2} h_1 h_2^{1/2}$. As above, the associated canonical transformation reduces $H$ to a free Boson system. Furthermore, as in \cite{CramerEisert} or \cite{CSE}, one could replace $\Z^d$ by more general graphs and the $\Lambda_L$ by an exhausting sequence of finite subgraphs.

We have chosen not to work in any of these more general settings, as our later applications to random oscillator systems will be stated in terms of the special case considered above. But generalizations of our results covering more general graphs or more general quadratic Hamiltonians are certainly possible.

\end{remark}

%
%
%

\section{Dynamical Localization} \label{sec:dlc}

The first set of main results of our work is aimed at establishing a form of dynamical localization for the Heisenberg evolution (\ref{eq:harmdyn}) of suitable observables under the Hamiltonian $H_L$. As discussed in the introduction, this will be done in the form of zero-velocity Lieb-Robinson bounds.

While in this and the following sections we work with finite systems $\Lambda_L$, the constants and decay rates in our bounds will be uniform in $L$. It is essentially for this reason that our proof of dynamical localization extends to infinite systems by taking the thermodynamic limit, as will be discussed in Section~\ref{sec:infinitevolume}.

The mechanism leading to dynamical localization in oscillator systems will be disorder, introduced by assuming that some of the system parameters are random variables. Exponential decay bounds will be proven for disorder averaged quantities. It is this disorder average which distinguishes our bounds from previously proven deterministic Lieb-Robinson bounds and allows to show that our bounds hold uniform in time, i.e.\ with group velocity zero.

Our general strategy for studying the Heisenberg dynamics of $H_L$ is to reduce it to the dynamics of the operator $h_L$ defined by (\ref{onepartham}), which will serve as an effective one-particle Hamiltonian for the many-body Hamiltonian $H_L$. We will state our main results, in this section as well as in Sections~\ref{sec:gsc} and \ref{sec:tsc}, as saying that localization properties of $h_L$ imply localization properties of $H_L$. In Section~\ref{sec:applications} we will provide regimes in which the required localization properties of $h_L$ can be verified.

In this section we will study the Heisenberg dynamics for two types of observables. Most natural may be to consider the dynamics of the position and momentum operators $q_x$ and $p_x$. We will postpone this to Section~\ref{subsec:comp+q}, while we will first study the dynamics of Weyl operators associated with positions and momenta. The main reason for this is that the Weyl operators are bounded (in fact, unitary) which will lead to a slightly more streamlined presentation of our results.

Throughout this section as well as in Sections~\ref{sec:gsc} and \ref{sec:tsc} we will assume {\bf A.1} and {\bf A.2} without further reference.

%
%
%

\subsection{Dynamical Localization for Weyl Operators} \label{subsec:comweyl}
In this section, we introduce the set of Weyl operators $W(f)$ generated by functions
$f : \Lambda_L \to \mathbb{C}$. In terms of these observables, we prove a strong
form of dynamical localization for certain random oscillator models, see Theorem~\ref{thm:sdlweylcom} and the
comments that follow.

To define Weyl operators, it is convenient to introduce annihilation and creation operators.
These are defined, for any $x \in \Lambda_L$,  by setting
\begin{equation} \label{eq:anncre}
a_x \, = \, \frac{1}{\sqrt{2}} \left( q_x \, + \, i p_x \right) \quad \mbox{and} \quad a^*_x \, = \, \frac{1}{\sqrt{2}} \left( q_x \, - \, i p_x \right) \, .
\end{equation}
Recalling (\ref{CCRsa}), it is easy to see that these operators satisfy the CCR, i.e.,
\begin{equation} \label{eq:cras}
[a_x, a_y] = [a^*_x, a^*_y] = 0 \quad \mbox{and} \quad [a_x, a^*_y] = \delta_{x,y} I \, .
\end{equation}
More generally, for any $f : \Lambda_L \to \mathbb{C}$, one can set
\begin{equation} \label{eq:defafa*f}
a(f) \, = \, \sum_{x \in \Lambda_L} \overline{f(x)} \, a_x  \quad \mbox{and} \quad a^*(f) \, = \, \sum_{x \in \Lambda_L} f(x) \, a_x^*.
\end{equation}
It follows from (\ref{eq:cras}) that
\begin{equation} \label{eq:comrela}
\left[ a(f), a(g) \right] = \left[ a^*(f), a^*(g) \right] =0 \,  \quad \mbox{and} \quad  \left[ a(f), a^*(g) \right] = \, \langle f, g \rangle \, .
\end{equation}
Each $f$, as above, corresponds to a Weyl operator $W(f)$ defined by
\begin{equation} \label{eq:weylina}
W(f) \, = \, {\rm exp} \left[ \frac{i}{\sqrt{2}} \left( a(f) \, + \, a^*(f) \right) \right] = {\rm exp} \left[ i \sum_x \left( {\rm Re}[f(x)] q_x + {\rm Im}[f(x)] p_x \right) \right]\, ,
\end{equation}
where the second identity is (\ref{eq:atopq}) below. Clearly, $W(f)$ is a unitary operator on $\mathcal{H}_L$ with $W^*(f) = W(-f)$ and $W(0) = I$.
As is well-known, see e.g.\ \cite{BratRob}, Weyl operators satisfy the so-called Weyl relations, i.e.
for any such $f$ and $g$
\begin{equation} \label{eq:weylrel}
W(f+g) = e^{ \frac{i}{2} {\rm Im}[\langle f, \, g \rangle]}   W(f) \, W(g) \, .
\end{equation}
These Weyl operators depend implicitly on the volume $\Lambda_L$, but we
suppress this in our notation.

A convenient deterministic fact about these Weyl operators is that they are, as a set, invariant under the harmonic dynamics.
For completeness, we state this as a lemma.
\begin{lemma} \label{lem:invweyl} For each finite set $\Lambda_L$, let $\tau_t$ denote the corresponding harmonic dynamics.
For any $f : \Lambda_L \to \mathbb{C}$ and $t \in \mathbb{R}$,
\begin{equation} \label{eq:harmevoweyl}
\tau_t(W(f)) = W(f_t) \quad \mbox{where} \quad f_t =  V^{-1}e^{2i \gamma t} Vf
\end{equation}
with a real linear mapping $V$ defined by
\begin{equation} \label{eq:defV}
Vf = \gamma^{-1/2} \mathcal{O}^T \mu^{1/2}{\rm Re}[f] + i \gamma^{1/2} \mathcal{O}^T \mu^{-1/2}{\rm Im}[f] \, ,
\end{equation}
and $\gamma$, $\mathcal{O}$, and $\mu$ are as in Section~\ref{Sec:OscSystems}.
\end{lemma}
\begin{proof}
To prove this fact, we first express the Weyl operators in terms of the
operators $B_k$ that diagonalize the Hamiltonian, see (\ref{diagH}). Using vector notation, write $a = (a_x)$ and $a^* = (a_x^*)$ and observe that
\begin{eqnarray} \label{eq:atopq} \\
a(f) + a^*(f) \; = \; ( \overline{f}^T, f^T) \left( \begin{array}{c} a \nonumber \\ a^* \end{array} \right) & = & \frac{1}{\sqrt{2}} ( \overline{f}^T, f^T) \left( \begin{array}{cc} 1 & i  \\ 1  & - i  \end{array} \right)  \left( \begin{array}{c} q \\ p \end{array} \right) \nonumber \\ & = & \sqrt{2} ({\rm Re}[f]^T, {\rm Im}[f]^T) \left( \begin{array}{c} q \\ p \end{array} \right)\, \nonumber
\end{eqnarray}
Appealing to (\ref{eq:q+pvecB}), it follows that
\begin{eqnarray} \label{eq:astobs} \\
a(f) + a^*(f) & = & ({\rm Re}[f]^T, {\rm Im}[f]^T) \left( \begin{array}{cc} \mu^{1/2} \mathcal{O} \gamma^{-1/2} & 0 \nonumber \\ 0 & \mu^{-1/2} \mathcal{O} \gamma^{1/2} \end{array} \right)
\left( \begin{array}{cc}  1 &  1 \\ -i & i \end{array} \right) \left( \begin{array}{c} B \\ B^* \end{array} \right) \nonumber \\
& = & ( (\overline{Vf})^T, (Vf)^T)  \left( \begin{array}{c} B \\ B^* \end{array} \right) = B(Vf) + B^*(Vf) \nonumber
\end{eqnarray}
with a real-linear mapping $V$ given by (\ref{eq:defV}). Here, in analogy to (\ref{eq:defafa*f}), we have also introduced the notation
\begin{equation} \label{eq:defB+B*phi}
B( \phi) \, = \, \sum_k \overline{\phi(k)} \, B_k  \quad \mbox{and} \quad B^*(\phi) \, = \, \sum_k \phi(k) \, B_k^*.
\end{equation}
We have shown that
\begin{equation} \label{eq:weylinB}
W(f) \, = \, {\rm exp} \left[ \frac{i}{\sqrt{2}} \left( a(f) \, + \, a^*(f) \right) \right] \, = \, {\rm exp} \left[ \frac{i}{\sqrt{2}} \left( B(Vf) \, + \, B^*(Vf) \right) \right] \, .
\end{equation}
The claim in (\ref{eq:harmevoweyl}) now follows from (\ref{eq:bkevolv}).
\end{proof}

Before we state our first result, the following observations will be useful.
It is easy to check that the inverse of the mapping in (\ref{eq:defV}) is given by
\begin{equation}
V^{-1} \phi = \mu^{-1/2} \mathcal{O} \gamma^{1/2} {\rm Re}[ \phi] + i \mu^{1/2} \mathcal{O} \gamma^{-1/2} {\rm Im}[ \phi] \, ,
\end{equation}
in which case, one readily verifies that
\begin{equation} \label{eq:reft}
{\rm Re}[f_t] = \mu^{-1/2} \cos(2t h_L^{1/2}) \mu^{1/2} {\rm Re}[f] - \mu^{-1/2} h_L^{1/2} \sin(2t h_L^{1/2}) \mu^{-1/2} {\rm Im}[f]
\end{equation}
whereas
\begin{equation} \label{eq:imft}
{\rm Im}[f_t] = \mu^{1/2} \cos(2t h_L^{1/2}) \mu^{-1/2} {\rm Im}[f] + \mu^{1/2} h_L^{-1/2} \sin(2t h_L^{1/2}) \mu^{1/2} {\rm Re}[f] \, .
\end{equation}
Here, and below, functions of the operator $h_L$ are defined via the elementary functional calculus of self-adjoint operators in finite-dimensional Hilbert spaces, e.g.\ $\cos(2t h_L^{1/2}) = \mathcal{O} \cos(2t \gamma) \mathcal{O}^T$, and so on. Note that, due to (\ref{diagtildeh}), we have $\mathcal{O} \gamma \mathcal{O}^T = h_L^{1/2}$, which is the reason for the frequent appearance of $h_L^{1/2}$ in relations between the one-particle operator $h_L$ and the many-body operator $H_L$, here as well as below. In fact, it is more accurate to think of $h_L^{1/2}$ rather than of $h_L$ as the effective one-particle Hamiltonian associated with $H_L$.

The identity (\ref{eq:harmevoweyl}) in conjunction with (\ref{eq:reft}) and (\ref{eq:imft}) allows us to derive dynamical localization properties of $H_L$ from localization properties of $h_L$. All our considerations so far have been deterministic,  meaning that they hold for any choice of the parameters $k_x$, $m_x$ and $\lambda_{x,y}$ satisfying our basic assumptions {\bf A.1} and {\bf A.2}. For our results on localization we will think of these parameters  as random variables on a probability space $(\Omega, \PP)$, which satisfy {\bf A.1} and {\bf A.2} almost surely. By $\E(X)$ we denote the expectation (average) of a random variable $X$ on $\Omega$ with respect to $\PP$.

For the sake of definiteness, one may think of the special case where one or several of the three parameter sets are given by i.i.d.\ random variables. In fact, our applications in Section~\ref{sec:applications} will be for the case where the $k_x$ are i.i.d., while the other parameters will be kept deterministic (for example, constant). But our general results below would, at least in principle, allow for applications with other types of disorder.

The required one-particle localization property will generally be expressed in the following form:

\begin{definition} \label{def:singefcor}
We say that $h_L$ has exponentially decaying $(\alpha,r)$-eigenfunction correlators if there exist $C<\infty$ and $\mu>0$ such that
\begin{equation} \label{eq:dynlocgeneral}
\mathbb{E} \left( \sup_{|u|\leq 1} \left| \langle \delta_x, h_L^{\alpha} u(h) \delta_y \right|^r \right) \leq Ce^{- \mu|x-y|}
\end{equation}
for all positive integers $L$ and all $x,y \in \Lambda_L$. Here the supremum is taken over all functions $u:\R\to \C$ which satisfy the pointwise bound $|u|\le 1$.
\end{definition}

As discussed in the introduction and, more thoroughly in Appendix~\ref{app:Anderson}, eigenfunction correlators for $\alpha=0$ are a well-known object in the literature on Anderson localization. A new aspect of our work is that we will need them for other, in particular negative, values of $\alpha$.

Our first result for disordered oscillator systems establishes dynamical localization for the Heisenberg evolution of Weyl operators.

\begin{theorem} \label{thm:sdlweylcom} Assume that $h_L$ has exponentially decaying $(-1/2,r)$-eigenfunction correlators for some $r\in (0,1]$. Then there exist $C'$ and $\mu'$ for which
\begin{equation} \label{eq:avweylcombd}
\mathbb{E} \left( \sup_{t \in \mathbb{R}} \left\| \left[ \tau_t(W(f)), W(g) \right] \right\| \right) \leq C' \sum_{x,y} |f(x)|^r |g(y)|^r e^{- \mu' |x-y|}
\end{equation}
for any $L \geq 1$ and $f, g: \Lambda_L \to \mathbb{C}$.
\end{theorem}

\begin{proof}
Let $f$ and $g$ be as above. Using Lemma~\ref{lem:invweyl} and the Weyl relations (\ref{eq:weylrel}), it is easy to see that
\begin{equation}
\left[ \tau_t(W(f)), W(g) \right] = \left( e^{-i {\rm Im}[ \langle f_t, g \rangle]} -1 \right) W(g) W(f_t) \, .
\end{equation}
In this case, the norm bound
\begin{equation} \label{eq:commbd}
\left\| \left[ \tau_t(W(f)), W(g) \right]  \right\| \leq \min \left\{ 2, \left| {\rm Im}[ \langle f_t, g \rangle] \right| \right\} \,
\end{equation}
readily follows since the Weyl operators are unitary. Moreover, since
\begin{equation}
{\rm Im}[ \langle f_t, g \rangle ] = \langle {\rm Re}[f_t], {\rm Im}[g] \rangle - \langle {\rm Im}[f_t], {\rm Re}[g] \rangle \, ,
\end{equation}
a short calculation, using (\ref{eq:reft}) and (\ref{eq:imft}), shows that
\begin{eqnarray} \label{eq:longform} \\
\lefteqn{{\rm Im}[ \langle f_t, g \rangle ]} \nonumber \\
& = & \langle \mu^{1/2} {\rm Re}[f], h_L^{-1/2} \sin(2th_L^{1/2}) \mu^{1/2} {\rm Re}[g] \rangle + \langle \mu^{-1/2} {\rm Im}[f], \cos(2th_L^{1/2}) \mu^{1/2} {\rm Re}[g] \rangle \nonumber \\
& & \mbox{} + \langle \mu^{1/2} {\rm Re}[f], \cos(2th_L^{1/2}) \mu^{-1/2} {\rm Im}[g] \rangle - \langle \mu^{-1/2} {\rm Im}[f], h_L^{1/2} \sin(2th_L^{1/2}) \mu^{-1/2} {\rm Re}[g] \rangle \, . \nonumber
\end{eqnarray}
By {\bf A.1} the entries of the diagonal matrices $\mu^{1/2}$ and $\mu^{-1/2}$ satisfy almost sure $L$-independent constant bounds. As a consequence, almost surely,
\begin{eqnarray} \label{eq:weylcombd} \\
\sup_{t\in \R} \left\| \left[ \tau_t(W(f)), W(g) \right]  \right\| & \leq & \sup_{t\in\R} 2^{1-r} \left|  {\rm Im}[ \langle f_t, g \rangle]  \right|^r  \nonumber \\
& \leq &  C_1 \sum_{x,y} |f(x)|^r |g(y)|^r \sum_{\alpha \in \{ -1/2, 0, 1/2 \}} \sup_{|u| \leq 1} \left| \langle \delta_x, h_L^{\alpha} u(h) \delta_y \rangle \right|^r \, . \nonumber
\end{eqnarray}
The supremum over $t\in\R$ has been absorbed into the supremum over $|u|\le 1$. Among the three values of $\alpha$ on the right of (\ref{eq:weylcombd}), the term with $\alpha=-1/2$ is the most singular and crucial one. In fact, if $|u|\le 1$ and $\tilde{u}(s) := s^{1/2} u(s)$, then $|\tilde{u}| \le( \max \{\sigma(h_L)\})^{1/2}$ on the spectrum $\sigma(h_L)$ of $h_L$. From this it readily follows that
\begin{equation}
\sup_{|u|\le 1} | \langle \delta_x, u(h_L) \delta_y \rangle| \le C_2 \sup_{|u|\le 1} |\langle \delta_x, h_L^{-1/2} u(h) \delta_y \rangle|,
\end{equation}
where $C_2$ can be chosen as the square root of the constant on the right hand side of (\ref{eq:normboundh}). In a similar way, the $\alpha=1/2$ term in (\ref{eq:weylcombd}) can also be absorbed in the $\alpha=-1/2$ term. The claim in (\ref{eq:avweylcombd}) now follows from (\ref{eq:dynlocgeneral}) with $\alpha=-1/2$.
\end{proof}

Two remarks are in order to clarify the relation of our assumptions to the different $\alpha$-terms in (\ref{eq:weylcombd}):

\begin{remark} \label{rem3} As is clear from (\ref{eq:longform}), the most singular term $\alpha=-1/2$ in (\ref{eq:weylcombd}) is due to the real parts of $f$ and $g$, which, by (\ref{eq:weylina}) and (\ref{eq:atopq}), correspond to contributions by positions to the Weyl operators. If $f$ and/or $g$ are purely imaginary (corresponding to Weyl operators only involving momenta), then a bound of the form (\ref{eq:avweylcombd}) can be proven under (for the sake of applications) weaker assumptions than exponential decay of $(-1/2,r)$ correlators. For example, exponentially decaying $(1/2,r)$-eigenfunction correlators for some $r\in (0,1]$ suffice to conclude that
\begin{equation}
\mathbb{E} \left( \sup_{t\in\R} \| [ \tau_t(e^{ip_x}), e^{ip_y}]\| \right) \le C' e^{-\mu'|x-y|}.
\end{equation}
This non-symmetry of our results in the positions and momenta is, of course, due to our initial choice of the model (\ref{generalHam}), where interactions are introduced via the position operators.
\end{remark}

\begin{remark} \label{rem4} That in the above proof the $\alpha=0$ and $\alpha=1/2$ terms could be absorbed into the $\alpha=-1/2$ term was due to the fact that our assumptions give a norm bound on $h_L$ which is uniform in the disorder as well as in the size of the system, thus avoiding that the $\alpha=1/2$ term introduces a singularity at infinite energy. Similarly, we could avoid the zero-energy singularity arising from the $\alpha=-1/2$ term by, in addition to {\bf A.1}, requiring a lower bound $k_x \ge k_{\rm min}>0$, uniformly in $x$ and the disorder. This would show $h_L\ge k_{\rm min}/2$ and, via (\ref{onepartham}) and (\ref{eq:evalsH}), result in a uniform ground state gap for the oscillator system $H_L$ (with an explicit lower bound given by $(k_{\rm min}/m_{\rm max})^{1/2}$). In this case  all $(\alpha,r)$ correlators would be equivalent in the sense that their exponential decay would not depend on the choice of $\alpha$.
\end{remark}

The following corollary is an immediate consequence of (\ref{eq:avweylcombd}) in cases where $f$ and $g$ have disjoint support. It phrases our result in a form more similar to traditional Lieb-Robinson bounds and explains why we have referred to it in the introduction as a zero-velocity Lieb-Robinson bound.

\begin{corollary} \label{cor:sdlcomweyl} Fix two finite sets $X, Y \subset \mathbb{Z}^d$ with $X \cap Y = \emptyset$. Let $L \geq 1$ be sufficiently large
so that $X \cup Y \subset \Lambda_L$. If $h_L$ has exponentially decaying $(-1/2,r)$-correlators for some $r\in (0,1]$, then there
are $C' < \infty$ and $\mu'>0$ such that
\begin{equation} \label{eq:zerovelLR}
\mathbb{E} \left( \sup_{t \in \mathbb{R}} \left\| \left[ \tau_t(W(f)), W(g) \right] \right\| \right) \leq C' \| f \|_{r}^r \| g \|_{r}^r e^{- \mu' d(X,Y)} \,
\end{equation}
holds for all $f$ and $g$ supported in $X$ and $Y$ respectively. Here $d(X,Y) = \min_{x \in X, y \in Y}|x-y|$.
\end{corollary}

In \cite{NRSS} oscillator systems of the form (\ref{generalHam}) with constant parameters $m_x = m$, $k_x = k$, $\lambda_{x,x+e_j} = \lambda_j$, $j=1,\ldots,d$, were considered and, in particular, a deterministic LR-bound of the form
\begin{equation}
\left\| \left[ \tau_t(W(f)), W(g) \right] \right\| \le C \|f\|_{\infty} \|g\|_{\infty} \min\{|X|,|Y|\} e^{-\mu(d(X,Y)-v|t|)}
\end{equation}
was proven for this case, see Corollary~3.2 in \cite{NRSS}. The finite positive number $v$ is interpreted as a bound on the velocity of propagation of signals in the oscillator system. In the cases discussed in Section~\ref{sec:applications} where we can verify exponential decay of $(-1/2,r)$-eigenfunction correlators for $h_L$ and after disorder averaging, our result shows that this velocity is zero for the random oscillator systems considered here.

More general deterministic oscillator systems (allowing for variable coefficients and more general lattices) were subsequently considered in \cite{CSE}, again establishing a ``finite speed of light'' in the system. We stress here that there were no previous results establishing dynamical localization of oscillator systems in the sense of a zero-velocity Lieb-Robinson bound such as (\ref{eq:zerovelLR}).

In addition to the dynamics of Weyl operators, the papers \cite{NRSS} and \cite{CSE} also have results for the dynamics of positions and momenta. The effects of disorder on the latter are the topic of the next subsection.

%
%
%

\subsection{Dynamical Localization for Positions and Momenta} \label{subsec:comp+q}
Theorem~\ref{thm:sdlweylcom} was concerned with the dynamics of Weyl operators. We will now consider an analogue for the dynamics of position and momentum operators. Due to the unboundedness of these operators,
the results are a bit different. We comment on these differences in this section.

We begin with a well-known result which we state as a lemma and prove for completeness.
For any $x, y \in \Lambda_L$ and $t \in \mathbb{R}$ set
\begin{equation}
A_{x,y}(t) = -i \left( \begin{array}{cc} \left[ \tau_t(q_x), q_y \right] & \left[ \tau_t(q_x), p_y \right]  \\
\left[ \tau_t(p_x), q_y \right]  & \left[ \tau_t(p_x), p_y \right]   \end{array} \right)
\end{equation}
where $\tau_t$ is the harmonic dynamics, see (\ref{eq:harmdyn}) above. Our normalization is such that
\begin{equation}
A_{x,y}(0) = \delta_{x,y} \left( \begin{array}{cc} 0 & I \\ -I & 0 \end{array} \right) \, .
\end{equation}

\begin{lemma} \label{lem:commat} For any $x, y \in \Lambda_L$ and each $t \in \mathbb{R}$, one has that
\begin{equation} \label{eq:comp+qatt}
A_{x,y}(t) = \left( \begin{array}{cc} - \langle \mu^{1/2} \delta_x, h_L^{-1/2} \sin(2 t h_L^{1/2}) \mu^{1/2} \delta_y \rangle I &  \langle \mu^{1/2} \delta_x, \cos(2 t h_L^{1/2}) \mu^{-1/2} \delta_y \rangle I \\ - \langle \mu^{-1/2} \delta_x, \cos(2 t h_L^{1/2}) \mu^{1/2} \delta_y \rangle I & - \langle \mu^{-1/2} \delta_x, h_L^{1/2} \sin(2 t h_L^{1/2}) \mu^{-1/2} \delta_y \rangle I \end{array} \right)
\end{equation}
where $\mu$ and $h_L$ are as in Section~\ref{Sec:OscSystems}.
\end{lemma}

\begin{proof}
The commutators appearing in $A_{x,y}(t)$ are entries in the larger matrix
\begin{equation} \label{eq:comm}
\left( \begin{array}{c} \tau_t(q) \\ \tau_t(p) \end{array} \right) \left( q^T, p^T \right) - \left( \left( \begin{array}{c} q \\ p \end{array} \right) \left( \tau_t(q)^T, \tau_t(p)^T \right) \right)^T
\end{equation}
Note that the CCR for the $B_k$'s can be expressed as
\begin{equation} \label{eq:CCRmat}
\left( \begin{array}{c} B \\ B^* \end{array} \right) \left( B^T, (B^*)^T \right) - \left( \left( \begin{array}{c} B \\ B^* \end{array} \right) \left( B^T, (B^*)^T \right) \right)^T = \left( \begin{array}{cc}  0 & I \\ -I & 0 \end{array} \right) \, .
\end{equation}
Using (\ref{eq:CCRmat}), a short calculation shows that the matrix in (\ref{eq:comm}) simplifies to
\begin{eqnarray} \label{eq:shortcalc} \\
& -i \left( \begin{array}{cc} \mu^{1/2} \mathcal{O} \gamma^{-1/2} & 0 \\ 0 & \mu^{-1/2} \mathcal{O} \gamma^{1/2} \end{array} \right) \left(  \begin{array}{cc} \sin(2 \gamma t) & - \cos(2 \gamma t) \\ \cos(2 \gamma t) & \sin( 2 \gamma t) \end{array} \right) \left(  \begin{array}{cc} \gamma^{-1/2} \mathcal{O}^T \mu^{1/2} & 0 \\ 0 & \gamma^{1/2} \mathcal{O}^T \mu^{-1/2} \end{array} \right)  \nonumber \\
& = -i \left( \begin{array}{cc} \mu^{1/2} \mathcal{O} \gamma^{-1} \sin( 2 \gamma t) \mathcal{O}^T \mu^{1/2} & - \mu^{1/2} \mathcal{O} \cos( 2 \gamma t) \mathcal{O}^T \mu^{-1/2} \\
\mu^{-1/2} \mathcal{O} \cos( 2 \gamma t) \mathcal{O}^T \mu^{1/2} &  \mu^{-1/2} \mathcal{O} \gamma \sin( 2 \gamma t) \mathcal{O}^T \mu^{-1/2} \end{array} \right) \nonumber
\end{eqnarray}
The claim in (\ref{eq:comp+qatt}) now follows by taking the appropriate matrix entries.
\end{proof}

In applying Lemma~\ref{lem:commat} to disordered oscillator systems, unlike with the Weyl operators, the upper left entries of (\ref{eq:comp+qatt}), corresponding to the position-position case, do not satisfy a norm bound which is uniform in the system size and the disorder (and thus we can not use an argument as in the first line of (\ref{eq:weylcombd})). This is due to the appearance of the factor $h_L^{-1/2}$, which can have arbitrarily large norm as $h_L$ may have eigenvalues arbitrarily close to zero. It is for this reason that the following result is stated in two parts, with an additional fractional power included in the commutator bound for the position-position case. Its proof follows immediately from Lemma~\ref{lem:commat}. For the sake of simplicity we do not state a third separate result for the momentum-momentum case, where the assumption could be weakened further by assuming exponential decay of $(1/2,1)$-eigenfunction correlators.

\begin{theorem} \label{thm:sdlp+qcom}

(a) If $h_L$ has exponentially decaying $(-1/2,r)$-eigenfunction correlators for some $r\in (0,1]$, then there exist $C'<\infty$ and $\mu'>0$ for which
\begin{equation} \label{eq:avp+qcombd}
\mathbb{E} \left( \sup_{t \in \mathbb{R}} \left\| \left[ \tau_t(q_x), q_y \right] \right\|^r \right) \leq C' e^{- \mu' |x-y|}
\end{equation}
for all $L$ and $x,y \in \Lambda_L$.

(b) If $h_L$ has exponentially decaying $(0,1)$-eigenfunction correlators, then there exist $C'<\infty$ and $\mu'>0$ for which
\begin{equation} \label{eq:avp+qcombd2}
\mathbb{E} \left( \sup_{t \in \mathbb{R}} \max \left\{ \| [ \tau_t(p_x), q_y ] \|, \| [ \tau_t(q_x), p_y ] \|, \| [ \tau_t(p_x), p_y ] \|  \right\} \right) \leq C' e^{- \mu' |x-y|}
\end{equation}
for all $L$ and $x,y \in \Lambda_L$.

\end{theorem}

We conclude this section by noting that it is tempting to try to derive Theorem~\ref{thm:sdlp+qcom} as a consequence of Theorem~\ref{thm:sdlweylcom}, at least if the assumption on eigenfunction correlators holds with $r=1$. Considering only position operators, (\ref{eq:avweylcombd}) would then imply that
\begin{equation} \label{eq:weyltopos}
\E \left( \sup_{t\in\R} \left\| \left[ \tau_t\left(\frac{e^{i\varepsilon q_x}-I}{i\varepsilon}\right), \frac{e^{i\varepsilon q_y}-I}{i\varepsilon} \right] \right\| \right) \le C' e^{-\mu'|x-y|}
\end{equation}
for every $\varepsilon>0$. Formally, this yields (\ref{eq:avp+qcombd}) as $\varepsilon\to 0$. However, as the position operators are unbounded, it is not obvious how to rigorously justify this limit.

%
%
%

\section{Ground State Correlations} \label{sec:gsc}

Localization near the bottom of the spectrum is sometimes described as the existence of a
``mobility gap'', which does not mean that there is an actual gap in the spectrum above the
ground state. In this section we show that in the models we consider a mobility gap implies
exponential decay of spatial correlations in the ground state.

As is discussed in Section~\ref{Sec:OscSystems} after (\ref{diagH}), the oscillator hamiltonians $H_L$ each have
a unique, normalized ground state $\Omega_L$ which is characterized by $B_k \Omega_L = 0$ for all $k = 1, 2, \ldots, | \Lambda_L|$.
To fix notation, for any linear operator $A$ in $\mathcal{H}_L$ such that $\Omega_L \in D(A)$, the domain of $A$, let us denote by
\begin{equation}
\langle A \rangle = \langle \Omega_L, A \Omega_L \rangle
\end{equation}
the ground state expectation of $A$. In this section, we will investigate static ground state correlations $\langle AB\rangle - \langle A \rangle \langle B \rangle$ as well as dynamic ground state correlations $\langle \tau_t(A)B\rangle - \langle \tau_t(A) \rangle \langle B \rangle =  \langle \tau_t(A)B\rangle - \langle A \rangle \langle B \rangle$. As in the previous section, we have results for the case where $A$ and $B$ are Weyl operators as well as for the case of positions and momenta.

%
%
%

\subsection{Correlations of Weyl Operators} \label{sec:gscorweyl}

We start with ground state correlations of Weyl operators.
As is well-known (or follows by taking the $\beta\to\infty$ limit in (\ref{eq:tscorweyl}) below), for any $f: \Lambda_L \to \mathbb{C}$
\begin{equation} \label{eq:gseweyl}
\langle W(f) \rangle = e^{- \frac{1}{4}\| Vf \|^2} \,
\end{equation}
where $V$ is the real-linear mapping given by (\ref{eq:defV}). Let us denote the dynamic ground state correlations of Weyl operators $W(f)$ and $W(g)$ by
\begin{equation} \label{eq:staticdef}
C(f,g;t) = \langle \tau_t(W(f)) W(g) \rangle - \langle W(f) \rangle \langle W(g) \rangle \, .
\end{equation}

In this context, our result is the following.

\begin{theorem} \label{thm:sdlweylcor} If $h_L$ has exponentially decaying $(-1/2,r)$-eigenfunction correlators for some $r\in (0,1]$, then there exist $C'< \infty$ and $\mu' >0$ such that
\begin{equation} \label{eq:avweylcorbd}
\mathbb{E} \left( \sup_{t \in \mathbb{R}} \left| C(f,g;t) \right| \right) \leq C' \sum_{x,y} |f(x)|^r |g(y)|^r e^{- \mu' |x-y|}
\end{equation}
for any $f, g \in \ell^2(\Lambda_L)$.
\end{theorem}

\begin{proof}
Using Lemma~\ref{lem:invweyl} and the Weyl relations (\ref{eq:weylrel}),
one sees that
\begin{eqnarray}
 \langle \tau_t(W(f)) W(g) \rangle & = &  \langle W(f_t) W(g) \rangle \\
 & = & e^{- \frac{i}{2} {\rm Im}[ \langle f_t, g \rangle]} \langle W(f_t+g) \rangle \nonumber \\
  & = & e^{- \frac{i}{2} {\rm Im}[ \langle f_t, g \rangle]} e^{-\frac{1}{4}\|V(f_t+g) \|^2} \, \nonumber
\end{eqnarray}
where we also used (\ref{eq:gseweyl}). Moreover, one has that
\begin{equation}
\| V(f_t+g) \|^2 = \| Vf_t \|^2 + \| Vg \|^2 + 2{\rm Re}[ \langle Vf_t, Vg \rangle ] \, .
\end{equation}
We have shown
\begin{equation}
C(f,g;t) = \left( e^{- \frac{i}{2} {\rm Im}[ \langle f_t, g \rangle]} e^{- \frac{1}{2} {\rm Re}[ \langle Vf_t, Vg \rangle]} -1 \right) e^{- \frac{1}{4} \left( \| Vf \|^2 + \|Vg \|^2 \right)} \, .
\end{equation}

Now for any real $a$ and $b$
\begin{equation}
e^{ia+b} - 1 = e^b(e^{ia} -1) + e^b -1  \quad \mbox{and so} \quad |e^{ia+b} -1| \leq (|a|+|b|)e^{|b|} \, .
\end{equation}
In this case,
\begin{equation}
|C(f,g;t)| \leq \frac{1}{2} \left( |{\rm Im}[ \langle f_t, g \rangle]| + |{\rm Re}[ \langle Vf_t, Vg \rangle]| \right)
e^{\frac{1}{2} \left| {\rm Re}[ \langle Vf_t, Vg \rangle] \right|}e^{- \frac{1}{4} \left( \| Vf \|^2 + \|Vg \|^2 \right)} \, .
\end{equation}
Since it is also true that
\begin{equation}
 \left| {\rm Re}[ \langle Vf_t, Vg \rangle] \right| \leq \| V f_t \| \| Vg \| = \| Vf \| \| V g \| \leq \frac{1}{2} \left( \|Vf \|^2 + \|V g \|^2 \right) \, ,
\end{equation}
the bound
\begin{equation}
|C(f,g;t)| \leq \frac{1}{2} \left( |{\rm Im}[ \langle f_t, g \rangle]| + |{\rm Re}[ \langle Vf_t, Vg \rangle]| \right)  \,
\end{equation}
follows.

The first term above is identical to the term appearing in (\ref{eq:commbd}), and we estimate it as before by using (\ref{eq:longform}).

The second term involves
\begin{eqnarray} \label{eq:longform2} \\
\lefteqn{{\rm Re}\left[ \langle Vf_t, Vg \rangle \right]} \nonumber \\ & = & \left\langle {\rm Re}[e^{2i \gamma t}Vf], {\rm Re}[Vg] \right\rangle + \left\langle {\rm Im}[e^{2i \gamma t}Vf], {\rm Im}[Vg] \right\rangle \nonumber \\
& = & \left\langle \cos(2 \gamma t) {\rm Re}[Vf], {\rm Re}[Vg] \right\rangle - \left\langle \sin(2 \gamma t) {\rm Im}[Vf], {\rm Re}[Vg] \right\rangle \nonumber \\
& \mbox{ } &  + \left\langle \sin(2 \gamma t) {\rm Re}[Vf], {\rm Im}[Vg] \right\rangle + \left\langle \cos(2 \gamma t) {\rm Im}[Vf], {\rm Im}[Vg] \right\rangle \nonumber \\
& = & \langle \mu^{1/2} {\rm Re}[f], h_L^{-1/2} \cos(2th_L^{1/2}) \mu^{1/2} {\rm Re}[g] \rangle - \langle \mu^{-1/2} {\rm Im}[f], \sin(2th_L^{1/2}) \mu^{1/2} {\rm Re}[g] \rangle \nonumber \\
& & \mbox{}  + \langle \mu^{1/2} {\rm Re}[f], \sin(2th_L^{1/2}) \mu^{-1/2} {\rm Im}[g] \rangle  + \langle \mu^{-1/2} {\rm Im}[f], h_L^{1/2} \cos(2th_L^{1/2}) \mu^{-1/2} {\rm Im}[g] \rangle \nonumber
\end{eqnarray}
which is very similar to (\ref{eq:longform}), the main distinction being that the role of sines and cosines is reversed. As $|C(f,g;t)|$ is uniformly bounded by $2$, we can argue as in (\ref{eq:weylcombd}) to introduce a fractional power $r\in (0,1]$. As in the proof of Theorem~\ref{thm:sdlweylcom}, the bound (\ref{eq:avweylcorbd})  now follows from the assumption on decay of eigenfunction correlators.

\end{proof}

For the averaged static correlations of Weyl operators, e.g.\ $t=0$ in (\ref{eq:staticdef}), one extracts the following explicit bound from the proof of Theorem~\ref{thm:sdlweylcor}.

\begin{corollary} \label{cor:staticWeylcor}
The static correlations of Weyl operators satisfy the bound
\begin{eqnarray} \label{eq:staticWeylcor}
\lefteqn{\E \left( |C(f,g;0)| \right)} \\
& \le & \frac{1}{2} |{\rm Im}[\langle f, g \rangle]| + C \sum_{x,y} |f(x)| |g(y)| \left( \E(|\langle \delta_x, h_L^{1/2} \delta_y \rangle|) + \E(|\langle \delta_x, h_L^{-1/2} \delta_y \rangle|) \right) \nonumber
\end{eqnarray}
for all $L$ and $f, g: \Lambda_L \to \C$. Here $C= \frac{1}{2} \max \{m_{\rm max}, 1/m_{\rm min} \}$ for the constants from {\bf A.1}.
\end{corollary}

This will allow to get exponential decay bounds on static correlations for a wider range of applications than on dynamic correlations. In particular, exponential decay of the terms $\E(|\langle \delta_x, h_L^{1/2} \delta_y \rangle|)$ and $\E(|\langle \delta_x, h_L^{-1/2} \delta_y \rangle|)$ will only require that the effective one-particle Hamiltonian $h_L$ is localized near energy $E=0$, while exponential decay of eigenfunction correlators (\ref{eq:dynlocgeneral}) requires localization of $h_L$ at {\it all} energies, see Section~\ref{sec:applications} and Appendix~\ref{app:Anderson}.

%
%
%

\subsection{Correlations of Positions and Momenta} \label{sec:gscorqp}

We next prove localization-type results for the ground state
correlations associated with the position and momentum operators.

Let $\mathcal{O}_j(x)$ be the matrix elements of the orthogonal matrix $\mathcal{O}$ from (\ref{diagtildeh}) with respect to the canonical bases in $\ell^2(\Lambda_L)$ and $\ell^2(\{1,\ldots, |\Lambda_L|\})$. Then the components of (\ref{eq:q+pvecB}) are given by
\begin{equation} \label{eq:qatx2}
q_x = \frac{1}{2} m_x^{-1/2}  \sum_k  \mathcal{O}_k(x) \gamma_k^{-1/2}  \left(B_k + B_k^* \right)
\end{equation}
and
\begin{equation} \label{eq:patx2}
p_x = i m_x^{1/2} \sum_k  \mathcal{O}_k(x) \gamma_k^{1/2}  \left(-B_k + B_k^* \right) \, .
\end{equation}
for each $x \in \Lambda_L$. Using these relations and the fact that $B_k \Omega_L =0$ for all $k$, we find that
\begin{equation} \label{eq:qpexpzero}
\langle p_x \rangle = \langle q_x \rangle =0 \quad \mbox{for all } x \in \Lambda_L \, .
\end{equation}

For the Heisenberg evolution of the positions and momenta we have from (\ref{eq:matp+qtev}),
\begin{equation} \label{eq:qtev}
\tau_t(q_x) = \frac{1}{2} m_x^{-1/2} \sum_k  \mathcal{O}_k(x) \gamma_k^{-1/2}  \left(e^{-2i \gamma_k t} B_k + e^{2i \gamma_k t} B_k^* \right)
\end{equation}
and
\begin{equation} \label{eq:ptev}
\tau_t(p_x) = i m_x^{1/2}  \sum_k  \mathcal{O}_k(x) \gamma_k^{1/2}  \left(-e^{-2i \gamma_kt} B_k + e^{2i \gamma_kt} B_k^* \right) \, .
\end{equation}
For given $x$ and $y$ in $\Lambda_L$, using (\ref{eq:qpexpzero}), we can express the dynamic ground state correlations of positions and momenta at $x$ and $y$, respectively, in terms of the effective one-particle Hamiltonian $h_L$ as follows:

\begin{lemma} \label{lem:gscor}
For any $x,y \in \Lambda_L$ and $t \in \mathbb{R}$,
\begin{eqnarray} \label{eq:gscormatrix}
\langle \tau_t(q_x) q_y \rangle & = & \frac{1}{4} (m_x m_y)^{-1/2} \langle \delta_x,  h_L^{-1/2} e^{-2it h_L^{1/2}} \delta_y \rangle\, , \\
\langle \tau_t(q_x) p_y \rangle & = &
\frac{i}{2} (m_y/m_x)^{1/2} \langle \delta_x,  e^{-2it h_L^{1/2}} \delta_y \rangle \, , \nonumber \\
\langle \tau_t(p_x) q_y \rangle & = &
-\frac{i}{2} (m_x/m_y)^{1/2} \langle \delta_x,  e^{-2it h_L^{1/2}} \delta_y \rangle \, , \nonumber \\
\langle \tau_t(p_x) p_y \rangle & = &
(m_x m_y)^{1/2} \langle \delta_x,  h_L^{1/2} e^{-2it h_L^{1/2}} \delta_y \rangle\, . \nonumber
\end{eqnarray}
\end{lemma}
\begin{proof}
We calculate the first identity. The rest are done similarly. Combining (\ref{eq:qatx2}) with (\ref{eq:qtev}) and using that the $B_k$ annihilate the ground state and that $B_k^* \Omega_L$ is normalized by (\ref{eq:evecs}), one gets
\begin{eqnarray}
\langle \tau_t(q_x) q_y \rangle & = & \frac{1}{4} (m_x m_y)^{-1/2} \sum_{k, k'} \mathcal{O}_k(x) \mathcal{O}_{k'}(y) \gamma_k^{-1/2} \gamma_{k'}^{-1/2} \times \\
& \mbox{ } & \quad \times \left\langle \left( e^{-2i \gamma_k t}B_k + e^{2i \gamma_kt}B_k^* \right) \left(B_{k'} + B_{k'}^* \right) \right\rangle  \nonumber \\
& = & \frac{1}{4} (m_x m_y)^{-1/2} \sum_{k, k'} \mathcal{O}_k(x) \mathcal{O}_{k'}(y) \gamma_k^{-1/2} \gamma_{k'}^{-1/2} e^{-2i \gamma_kt} \langle B_k B_{k'}^* \rangle \nonumber \\
& = & \frac{1}{4} (m_x m_y)^{-1/2} \sum_k \mathcal{O}_k(x) \mathcal{O}_{k}(y) \gamma_k^{-1} e^{-2i \gamma_kt} \, \nonumber \\
& = & \frac{1}{4} (m_x m_y)^{-1/2} \langle \delta_x, \mathcal{O} \gamma^{-1} e^{-2i\gamma t} \mathcal{O}^T \delta_y \rangle, \nonumber
\end{eqnarray}
which gives (\ref{eq:gscormatrix}).
\end{proof}

An immediate consequence is

\begin{theorem} \label{thm:sdlp+qcor} (a) If $h_L$ has exponentially decaying $(-1/2,r)$-eigenfunction correlators for some $r\in (0,1]$, then there exist $C'<\infty$ and $\mu'>0$ such that
\begin{equation} \label{eq:dyncorgsqq}
\mathbb{E} \left( \sup_{t \in \mathbb{R}} | \langle \tau_t(q_x) q_y \rangle |^r \right) \leq C' e^{- \mu' |x-y|}
\end{equation}
for all $L$ and $x,y \in \Lambda_L$.

(b) If $h_L$ has exponentially decaying $(0,1)$-eigenfunction correlators, then there exist $C'<\infty$ and $\mu'>0$ such that
\begin{equation} \label{eq:dyncorgsp}
\mathbb{E} \left( \sup_{t \in \mathbb{R}} \max \{ | \langle \tau_t(p_x) q_y \rangle |, | \langle \tau_t(q_x) p_y \rangle |, | \langle \tau_t(p_x) p_y \rangle | \} \right) \leq C' e^{- \mu' |x-y|}
\end{equation}
for all $L$ and $x,y \in \Lambda_L$.
\end{theorem}

As for the Weyl operators in Section~\ref{sec:gscorweyl}, we also get simple explicit bounds for the static correlations of positions and momenta:

\begin{corollary} \label{cor:staticqpcor}
We have $\langle q_x p_y \rangle = -\langle p_x q_y \rangle = \frac{i}{2} \delta_{x,y}$ as well as
\begin{equation}
\E ( | \langle q_x q_y \rangle |) \le \frac{1}{4m_{\rm min}} \E (|\langle \delta_x, h_L^{-1/2} \delta_y \rangle|)
\end{equation}
and
\begin{equation}
\E(|\langle p_x p_y \rangle|) \le m_{\rm max} \E(|\langle \delta_x, h_L^{1/2} \delta_y \rangle|).
\end{equation}
\end{corollary}

%
%
%

\section{Correlations for Thermal States} \label{sec:tsc}

On physical grounds (increasing entropy) one would expect the correlation length to decrease with
increasing temperature. We therefore expect that the conditions under which we can prove
exponential decay in the ground state, would also be sufficient to prove exponential decay at
positive temperatures. This is not entirely trivial however. In Section~\ref{sec:dlc} we proved
exponential decayin the ground state based on localization near the bottom of the spectrum.
At higher temperatures higher portions in the spectrum dominate the state and the corresponding
states may be delocalized. It is therefore not entirely obvious that exponential decay holds for all
temperatures. Nevertheless, in this section we prove exponential decay of spatial correlations
in the equilibrium states (also called thermal states) at any temperature.

Thermal states are defined as follows.
Fix $\beta \in (0, \infty)$. For any $A$ such that $Ae^{-\beta H_L}$ is trace class, set
\begin{equation} \label{eq:+state}
\langle A \rangle_{\beta} = \frac{ {\rm Tr} \left[ A e^{- \beta H_L} \right]}{ {\rm Tr} \left[ e^{- \beta H_L} \right]}
\end{equation}
to be the expected value of $A$ in a thermal state corresponding to a positive temperature inversely proportional to $\beta<\infty$. Note that, as for zero temperature, $\langle \tau_t(A) \rangle_{\beta} = \langle A \rangle_{\beta}$.

%
%
%

\subsection{Correlations of Weyl Operators} \label{sec:tscorweyl}

We start with the calculation of thermal state correlations of Weyl operators, which provides a positive temperature analogue of the results in Section~\ref{sec:gscorweyl}.

It is well-known, see Proposition~5.2.28 of \cite{BratRob} or Chapter~XII.12 of \cite{Messiah1999}, that
\begin{equation} \label{eq:tscorweyl}
\langle W(f) \rangle_{\beta} = e^{- \frac{1}{4} \left\| A_{\beta}^{1/2} Vf \right\|^2}
\end{equation}
where $V$ is as in (\ref{eq:defV}) and
\begin{equation}
A_{\beta} = (I+e^{-2\beta \gamma})(I-e^{-2\beta \gamma})^{-1} = \coth(\beta\gamma).
\end{equation}
Here $\gamma$ is the diagonal operator from (\ref{diagtildeh}).

The thermal correlation of the Weyl operators $W(f)$ and $W(g)$ is
\begin{equation}
C_{\beta}(f,g;t) = \langle \tau_t(W(f)) W(g) \rangle_{\beta} - \langle W(f) \rangle_{\beta} \langle W(g) \rangle_{\beta} \, .
\end{equation}
Moreover, the Weyl relations (\ref{eq:weylrel}), (\ref{eq:harmevoweyl}) and (\ref{eq:tscorweyl}) imply that
\begin{equation}
\langle \tau_t(W(f)) W(g) \rangle_{\beta} = e^{- \frac{i}{2} {\rm Im}[ \langle f_t, g \rangle]}
e^{- \frac{1}{4}\| A_{\beta}^{1/2} V(f_t+g) \|^2} \, .
\end{equation}
Like before, the relation
\begin{equation}
\| A_{\beta}^{1/2} V(f_t+g) \|^2 = \| A_{\beta}^{1/2} Vf_t \|^2 +\| A_{\beta}^{1/2} Vg \|^2 + 2 {\rm Re}[ \langle A_{\beta} Vf_t, Vg \rangle]
\end{equation}
holds and therefore we obtain
\begin{equation}
| C_{\beta}(f,g;t) | \leq \frac{1}{2} \left( |{\rm Im}[ \langle f_t, g \rangle]| + |{\rm Re}[ \langle A_{\beta} Vf_t, Vg \rangle]| \right)
\end{equation}
as in Section~\ref{sec:gscorweyl}. By a calculation generalizing (\ref{eq:longform2}) we get
\begin{eqnarray}
\lefteqn{{\rm Re}[\langle A_{\beta} Vf_t, Vg \rangle]}  \\
& = & \langle \mu^{1/2} {\rm Re}[f], \phi_{1,t}(h_L) \mu^{1/2} {\rm Re}[g] \rangle - \langle \mu^{-1/2} {\rm Im}[f], \phi_{2,t}(h_L) \mu^{1/2} {\rm Re}[g] \rangle \nonumber \\
& & \mbox{} + \langle \mu^{1/2} {\rm Re}[f], \phi_{2,t}(h_L) \mu^{-1/2} {\rm Im}[g] \rangle + \langle \mu^{-1/2} {\rm Im}[f], \phi_{3,t}(h_L) \mu^{-1/2} {\rm Im}[g] \rangle. \nonumber
\end{eqnarray}
Here we have set
\begin{equation} \label{eq:phifunctions}
\phi_{1,t}(s) = s^{-1/2} \coth(\beta s^{1/2}) \cos(2ts^{1/2}), \quad \phi_{2,t}(s) = \coth(\beta s^{1/2}) \sin(2ts^{1/2}), \quad \phi_{3,t}(s) = s \phi_{1,t}(s).
\end{equation}
The operator functions $\phi_{1,t}(h_L)$, $\phi_{2,t}(h_L)$ and $\phi_{3,t}(h_L)$ appearing here are more singular than the operator functions we have encountered earlier in (\ref{eq:longform}) and (\ref{eq:longform2}). In fact, due to the additional factor $\coth(\beta s^{1/2})$ which behaves like $1/(\beta s^{1/2})$ near $s=0$, we have $\phi_{1,t}(s) \sim 1/(\beta s)$ for the most singular function in (\ref{eq:phifunctions}).

The stronger singularities are the mathematically most interesting new feature appearing in our treatment of thermal state correlations, as opposed to the earlier results on Lieb-Robinson bounds and ground state correlations. In the applications in Section~\ref{sec:applications} below we will have to control more singular types of eigenfunction correlators of the effective one-particle Hamiltonian $h_L$ than what is needed in applications of our earlier results.

Under the assumption that the required more singular eigenfunction correlators are exponentially decaying, we thus get the following analogue of Theorem~\ref{thm:sdlweylcor} for the thermal state correlations of Weyl operators.

\begin{theorem} \label{thm:thermalweylcor}
If $h_L$ has exponentially decaying $(-1,r)$-eigenfunction correlators for some $r\in (0,1]$, then there exist $C'<\infty$ and $\mu'>0$ such that
\begin{equation} \label{eq:avweylcorbdT}
\mathbb{E} \left( \sup_{t \in \mathbb{R}} \left| C_{\beta}(f,g;t) \right| \right) \leq C' \sum_{x,y} |f(x)|^r |g(y)|^r e^{- \mu' |x-y|}
\end{equation}
for any $f, g \in \ell^2(\Lambda_L)$.
\end{theorem}

\begin{remark} In principle, one can derive Theorem~\ref{thm:sdlweylcor} from Theorem~\ref{thm:thermalweylcor} by taking the $\beta\to\infty$ limit.
This would, however, require us to work under the stronger assumption of $(-1,r)$-correlator decay instead of $(-1/2,r)$-correlator decay as in Theorem~\ref{thm:sdlweylcor}, which seems unnatural and weakens applicability. For this reason, we have separated our results into two sections.
\end{remark}

As in Section~\ref{sec:gscorweyl} we state an explicit bound for the static thermal correlations of Weyl operators, which will prove their exponential decay in a wider range of applications.

\begin{corollary} \label{cor:staticthermalWeylcor}
The static correlations of Weyl operators satisfy the bound
\begin{eqnarray} \label{eq:staticthermalWeylcor} \\
\lefteqn{\E \left( |C_{\beta}(f,g;0)|^{1/2} \right)} \nonumber \\ & \le & \frac{1}{2^{1/2}} |{\rm Im}[\langle f, g \rangle]|^{1/2} + C^{1/2} \sum_{x,y} |f(x)|^{1/2} |g(y)|^{1/2} \Big\{ \E(|\langle \delta_x, h_L^{1/2} \coth(\beta h_L^{1/2}) \delta_y \rangle|^{1/2})   \nonumber \\ & & \mbox{} \ + \E(|\langle \delta_x, h_L^{-1/2} \coth(\beta h_L^{1/2}) \delta_y \rangle|^{1/2}) \Big\} \nonumber
\end{eqnarray}
for all $L$ and $f, g: \Lambda_L \to \C$. Here $C= \frac{1}{2} \max \{m_{\rm max}, 1/m_{\rm min} \}$ for the constants from {\bf A.1}.
\end{corollary}

%
%
%

\subsection{Correlations of $p$'s and $q$'s} \label{sec:tscorp+q}

Finally, we consider thermal correlations of positions and momenta. For them we have the following deterministic facts.

\begin{lemma} \label{lem:tscor} Fix $\beta >0$.  One has that
\begin{equation} \label{eq:0expts}
\langle q_x \rangle_{\beta} = \langle p_x \rangle_{\beta} = 0 \quad \mbox{for all } x \in \Lambda_L \, .
\end{equation}
Moreover, for any $x,y \in \Lambda_L$ and $t \in \mathbb{R}$,
\begin{eqnarray} \label{eq:tspqcor}
\langle \tau_t(q_x) q_y \rangle_{\beta} & = & \frac{1}{4} (m_x m_y)^{-1/2} \langle \delta_x, \left\{ \phi_{1,t}(h_L) - i h_L^{-1/2} \sin(2th_L^{1/2}) \right\} \delta_y \rangle  \\
\langle \tau_t(q_x) p_y \rangle_{\beta} & = & \frac{i}{2} (m_y/m_x)^{1/2} \langle \delta_x, \left\{ \cos(2th_L^{1/2})-i \phi_{2,t}(h_L) \right\} \delta_y \rangle \nonumber \\
\langle \tau_t(p_x) q_y \rangle_{\beta} & = & -\frac{i}{2} (m_x/m_y)^{1/2} \langle \delta_x, \left\{ \cos(2th_L^{1/2}) -i \phi_{2,t}(h_L) \right\} \delta_y \rangle \nonumber \\
\langle \tau_t(p_x) p_y \rangle_{\beta} & = & (m_x m_y)^{1/2} \langle \delta_x, \left\{ \phi_{3,t}(h_L) - ih_L^{1/2} \sin(2th_L^{1/2}) \right\} \delta_y \rangle \nonumber \, ,
\end{eqnarray}
where we use the functions defined in (\ref{eq:phifunctions}).
\end{lemma}

\begin{proof} This is proven by calculations similar to those used to prove Lemma~\ref{lem:gscor}, this time using the following facts on the thermal state expectations of the bosonic creation and annihilation operators:
\begin{equation}
\langle B_k \rangle_{\beta} = \langle B_k^* \rangle_{\beta} = \langle B_k B_{k'} \rangle_{\beta} = \langle B_k^* B_{k'}^* \rangle = 0 \, ,
\end{equation}
\begin{equation}
\langle B_k B_{k'}^* \rangle_{\beta} = (1-e^{-2\beta \gamma_k})^{-1} \delta_{k,k'}, \quad \langle B_k^* B_{k'} \rangle_{\beta} = (e^{2\beta \gamma_k} -1)^{-1} \delta_{k,k'} \, ,
\end{equation}
see Proposition~5.2.28 of \cite{BratRob}.

\end{proof}

The crucial difference between (\ref{eq:gscormatrix}) and (\ref{eq:tspqcor}) is that in the latter the factor $\coth(\beta h_L^{1/2})$ in $\phi_{1,t}(h_L)$, $\phi_{2,t}(h_L)$ and $\phi_{3,t}(h_L)$ leads to a stronger zero-energy singularity, in each case corresponding to an extra factor $h_L^{-1/2}$. Thus in each of the two parts of the following thermal state analogue of Theorem~\ref{thm:sdlp+qcor} a correspondingly more singular eigenfunction correlator needs to be controlled in the assumption. Otherwise the claims follow immediately from Lemma~\ref{lem:tscor}.

\begin{theorem} \label{thm:thermalpqcor}
(a) If $h_L$ has exponentially decaying $(-1,r)$-eigenfunction correlators for some $r\in (0,1]$, then there exist $C'<\infty$ and $\mu'>0$ such that
\begin{equation} \label{eq:tsdyncorqq}
\mathbb{E} \left( \sup_{t \in \mathbb{R}} | \langle \tau_t(q_x) q_y \rangle_{\beta} |^r \right) \leq C' e^{- \mu' |x-y|}
\end{equation}
for all $L$ and $x,y \in \Lambda_L$.

(b) If $h_L$ has exponentially decaying $(-1/2,r)$-eigenfunction correlators for some $r\in (0,1]$, then there exist $C'<\infty$ and $\mu'>0$ such that
\begin{equation} \label{eq:tsdyncorp}
\mathbb{E} \left( \sup_{t \in \mathbb{R}} \max \{ | \langle \tau_t(p_x) q_y \rangle_{\beta}|^r, | \langle \tau_t(q_x) p_y \rangle_{\beta}|^r, | \langle \tau_t(p_x) p_y \rangle_{\beta}|^r \} \right) \leq C' e^{- \mu' |x-y|}
\end{equation}
for all $L$ and $x,y \in \Lambda_L$.
\end{theorem}

As before, for the static correlations we state an explicit bound tailored to our applications in Section~\ref{sec:applications}.

\begin{corollary} \label{cor:tsstaticqpcor}
We have $\langle q_x p_y \rangle_{\beta} = -\langle p_x q_y \rangle_{\beta} = \frac{i}{2} \delta_{x,y}$ as well as
\begin{equation}
\E ( | \langle q_x q_y \rangle_{\beta} |^{1/2}) \le \frac{1}{2 m_{\rm min}^{1/2}} \E (|\langle \delta_x, h_L^{-1/2} \coth(\beta h_L^{1/2}) \delta_y \rangle|^{1/2})
\end{equation}
and
\begin{equation}
\E(|\langle p_x p_y \rangle_{\beta}|) \le m_{\rm max} \E(|\langle \delta_x, h_L^{1/2} \coth(\beta h_L^{1/2}) \delta_y \rangle|).
\end{equation}
\end{corollary}

\section{Applications} \label{sec:applications}

%
%
%

Our results on disordered oscillator systems $H_L$ in Sections~\ref{sec:dlc}, \ref{sec:gsc} and \ref{sec:tsc}  were stated in terms of assumptions on the exponential decay of eigenfunction correlators (\ref{eq:dynlocgeneral}) of the associated one-particle Hamiltonian $h_L$. Thus applications of our results will consist in providing conditions on the choice of the random parameters $m_x$, $k_x$ and $\lambda_{x,y}$ under which decay of eigenfunction correlators can be verified.

The main application will be what we will refer to as the Anderson case, which will describe the following set of conditions:

\vspace{.2cm}

{\bf Assumption A.3:} Let $m_x = m \in (0,\infty)$ and $\lambda_{x,y} = \lambda \in (0,\infty)$ be constant, and $k_x$, $x\in \Z^d$, be i.i.d.\ random variables with absolutely continuous distribution given by a bounded density $\rho$ supported in $[0,k_{\rm max}]$, $0<k_{\rm max}<\infty$.

\vspace{.2cm}

With this choice, the operator $h_L$ in (\ref{onepartham}), up to constant rescaling, becomes the Anderson model discussed in Appendix~\ref{app:Anderson}. This makes a wealth of known localization properties available. In particular, exponential decay of $(0,1)$-eigenfunction correlators (see Definition~\ref{def:singefcor}) is well known in various regimes. In fact, it is in works on the Anderson model where the term {\it eigenfunction correlator} was first used in this context, e.g.\ \cite{Aizenman94}, \cite{ASFH} and also the recent review \cite{Stolz11}.

However, {\it singular} eigenfunction correlators ($\alpha<0$ in (\ref{eq:dynlocgeneral})) do not seem to have appeared in previous works. Establishing localization bounds for them, after reviewing known results, is the main content of Appendix~\ref{app:Anderson} below. What makes the extension of localization results for eigenfunction correlators to singular eigenfunction correlators non-trivial is the fact that in {\bf A.3} we allow the $k_x$ to take values arbitrarily close to zero. As discussed in Remark~\ref{rem4} above, this means that the operator $h_L^{-1/2}$ is not uniformly bounded in the system size and the disorder and thus eigenfunction correlators for different values of $\alpha$ are not equivalent.

Expressed in terms of the many-body Hamiltonian $H_L$ this means that in cases where exponential decay of eigenfunction correlators can be shown under Assumption {\bf A.3}, we get many-body localization properties (e.g.\ zero-velocity LR-bounds as well as exponential clustering for ground states and thermal states) in situations where $H_L$ does {\it not} satisfy a stable ground state gap condition.

We group our applications to the Anderson case into three theorems, corresponding to the three different regimes  described in Proposition~\ref{prop:FML} where localization is known. We start with the general case {\bf A.3}, without restriction of the dimension, where localization is known to hold near the bottom of the spectrum. This does not allow to prove exponential decay of eigenfunction correlators, as required by all of our results in Sections~\ref{sec:dlc}, \ref{sec:gsc} and \ref{sec:tsc}, that refer to dynamical quantities. However, localization of $h_L$ near the bottom of the spectrum suffices to show exponential decay of the static ($t=0$) ground state and thermal state correlations for Weyl operators as well as positions and momenta:

\begin{theorem} \label{thm:applgenAnderson}
Assume {\bf A.3} and fix $\beta \in (0,\infty)$. Then there exist constants $C<\infty$ and $\mu>0$ such that
\begin{equation}
\E \left( |C(f,g;0)| \right) \le C \sum_{x,y} |f(x)| |g(y)| e^{-\mu|x-y|}
\end{equation}
and
\begin{equation}
\E \left( |C_{\beta}(f,g;0)|^{1/2} \right) \le C \sum_{x,y} |f(x)|^{1/2} |g(y)|^{1/2} e^{-\mu|x-y|}
\end{equation}
for all $L$ and $f, g \in \ell^2(\Lambda_L)$ with disjoint support. Moreover, all the quantities
\begin{equation}
\E(|\langle q_x q_y \rangle |), \quad \E(|\langle p_x p_y\rangle|), \quad \E(|\langle q_x q_y \rangle_{\beta}|^{1/2}), \quad \E (| \langle p_x p_y\rangle_{\beta}|^{1/2})
\end{equation}
satisfy bounds of the form $Ce^{-\mu|x-y|}$ uniformly in $L$ and $x,y\in \Lambda_L$.
\end{theorem}

Note that, as indicated in Corollaries~\ref{cor:staticqpcor} and \ref{cor:tsstaticqpcor}, the operators $q_x p_y$ and $p_x q_y$ have trivial ground and thermal state correlations, which we don't include here.

\begin{proof}
These results follow by combining Corollaries~\ref{cor:staticWeylcor}, \ref{cor:staticqpcor}, \ref{cor:staticthermalWeylcor} and \ref{cor:tsstaticqpcor}, whose assumptions hold in the Anderson regime due to Propositions~\ref{prop:EFCbounds}(c) and \ref{prop:verysingEFC}(b).

More precisely, the bounds on ground state correlations follow from Proposition~\ref{prop:EFCbounds}(c), as the functions $\varphi(s)=s^{1/2}$ and $\varphi(s)=s^{-1/2}$ have the required analyticity properties. Also, as $\varphi(s) = \coth(\beta s^{1/2})$ satisfies $|\varphi(s)| \le C/(\beta s^{1/2})$ for $s$ near $0$ and has an analytic extension to the right half plane, the assumptions of Proposition~\ref{prop:verysingEFC}(b) are satisfied with the choice $\varphi_1(s)= s^{\pm 1/2}$ and $\varphi_2(s) = \coth(\beta s^{1/2})$, leading to the decay bounds for thermal state correlations.

\end{proof}

As reviewed in Appendix~\ref{app:Anderson}, the Anderson model is localized at {\it all} energies, independent of the dimension, if the disorder is sufficiently large. This can be expressed by assuming that the density $\rho$ of the distribution of the $k_x$ has sufficiently small $L^{\infty}$-norm $\|\rho\|_{\infty}$, reflecting the fact that the values of the random variables $k_x$ must be widely spread. A frequent way of referring to this in applications is by replacing the $k_x$ with $\lambda k_x$, assuming that the distribution of $k_x$ has a density $\rho$ as in {\bf A.3} and that the disorder parameter $\lambda>0$ is sufficiently large (the random variables $\lambda k_x$ have density $\rho(\cdot/\lambda)/\lambda$ which have small $L^{\infty}$-norm for large $\lambda$).

Applying Propositions~\ref{prop:EFCbounds} and \ref{prop:verysingEFC} to the large disorder Anderson model leads to exponential decay of $(-1/2,1)$ as well as $(-1,1/2)$ eigenfunction correlators. This makes all our results on dynamics of disordered oscillator systems applicable, which we summarize in the following theorem.

\begin{theorem} \label{thm:applargedisorder}
Assume {\bf A.3} with sufficiently large disorder, i.e.\ that $\|\rho\|_{\infty}$ is sufficiently small. Then, in addition to the results of Theorem~\ref{thm:applgenAnderson}, we have the following dynamical localization properties of Lieb-Robinson commutators:

There are constants $C'<\infty$ and $\mu'>0$ such that
\begin{equation} \label{eq:applargedisorderWeyl}
\mathbb{E} \left( \sup_{t \in \mathbb{R}} \left\| \left[ \tau_t(W(f)), W(g) \right] \right\| \right) \leq C' \sum_{x,y} |f(x)| |g(y)| e^{- \mu' |x-y|}
\end{equation}
for all $L$ and $f, g:\Lambda_L \to \C$, as well as
\begin{equation} \label{eq:applargedisorderqp}
\E \left( \sup_{t\in \R} \| [\tau_t(c_x), d_y]\| \right) \le C'e^{-\mu'|x-y|}
\end{equation}
for all $L$ and $x,y \in \Lambda_L$, where $c_x \in \{q_x, p_x\}$ and $d_x \in \{q_y,p_y\}$.

Moreover, the bounds on dynamic ground state correlations (\ref{eq:avweylcorbd}), (\ref{eq:dyncorgsqq}) and (\ref{eq:dyncorgsp}) hold with $r=1$. The bounds on dynamic thermal state correlations (\ref{eq:avweylcorbdT}) and (\ref{eq:tsdyncorqq}) hold with $r=1/2$ and (\ref{eq:tsdyncorp}) holds with $r=1$.

\end{theorem}

\begin{proof} In the large disorder regime we can combine Proposition~\ref{prop:FML}(b) with Proposition~\ref{prop:EFCbounds}(b), choosing $E_0=\infty$ and $\alpha=-1/2$, to show that $h$ has exponentially decaying $(-1/2,1)$-eigenfunction correlators. Note here that the condition $\alpha > -s/(2-s)$ of Proposition~\ref{prop:EFCbounds}(b) is satisfied if $s>2/3$ is chosen (which is possible as Proposition~\ref{prop:FML}(b) allows for $s$ arbitrarily close to $1$).

That in the large disorder regime $h_L$ has exponentially decaying $(-1,1/2)$-eigenfunction correlators follows from Proposition~\ref{prop:verysingEFC}.

The theorems in Sections~\ref{sec:dlc}, \ref{sec:gsc} and \ref{sec:tsc} now yield all the claimed bounds.
\end{proof}

Another case in which localization can be proven at {\it all} energies is the one-dimensional regime, where no assumption on the disorder strength is required. The mechanisms behind one-dimensional localization are much more subtle than how localization follows (essentially by brute force) in the multi-dimensional large disorder regime. This leads to certain differences in the mathematical characteristics of localization as summarized in Appendix~\ref{app:Anderson}. As a consequence we observe some slight changes in the consequences for disordered oscillator systems, essentially in the form of ``regularizing exponents''. For simplicity, we state them only for our results on Lieb-Robinson commutators in the next result (which should be compared with (\ref{eq:applargedisorderWeyl}) and (\ref{eq:applargedisorderqp}) above) . Similar changes appear in some of the bounds on dynamical ground state and thermal state correlations.

\begin{theorem} \label{thm:apponedim}
Assume {\bf A.3} and $d=1$. Then there exist constants $C'<\infty$ and $\mu'>0$ such that
\begin{equation} \label{eq:app1DWeyl}
\E\left( \sup_{t\in \R} \|[\tau_t(W(f)), W(g)]\| \right) \le C' \sum_{x,y} |f(x)|^{1/2} |g(y)|^{1/2} e^{- \mu' |x-y|}
\end{equation}
for all $L$ and $f, g:\Lambda_L \to \C$, as well as
\begin{equation} \label{eq:app1Dqp}
\E \left( \sup_{t\in \R} \max \left\{ \| [\tau_t(q_x), q_y]\|^{1/2},  \| [\tau_t(p_x), q_y]\|,  \| [\tau_t(q_x), p_y]\|,  \| [\tau_t(p_x), p_y]\| \right\} \right) \le C'e^{-\mu'|x-y|}
\end{equation}
for all $L$ and $x,y \in \Lambda_L$.

\end{theorem}

\begin{proof} The changes and slightly weaker results as compared to Theorem~\ref{thm:applargedisorder} are due to the fact that we now have to use part (c) of Proposition~\ref{prop:FML} which establishes localization for $s$-moments only for sufficiently small $s>0$. This rules out an application of Proposition~\ref{prop:EFCbounds}(b) with $\alpha=-1/2$.  Instead we use Proposition~\ref{prop:verysingEFC}(a), which gives exponential decay of $(-1/2,1/2)$-eigenfunction correlators. This yields (\ref{eq:app1DWeyl}) and the claimed bound on the first term on the left hand side of (\ref{eq:app1Dqp}).

For the other terms we can use Theorem~\ref{thm:sdlp+qcom}(b), where exponential decay of $(0,1)$-correlators follows from Proposition~\ref{prop:EFCbounds}(a) (which applies with arbitrarily small $s$). Without having stated it, we note that if either $f$ and/or $g$ are purely imaginary, then one could replace (\ref{eq:app1DWeyl}) by (\ref{eq:applargedisorderWeyl}), see Remark~\ref{rem3} above.

\end{proof}

We conclude this section with some remarks about possible applications of our results in Sections~\ref{sec:dlc}, \ref{sec:gsc} and \ref{sec:tsc} to types of disorder other than the Anderson case discussed above, i.e.\ for other choices of the parameters $m_x$, $k_x$ and $\lambda_{x,y}$ of the oscillator system.

\vspace{.2cm}

(i) Many (likely all) of our results will generalize to the case where the masses $m_x$ and couplings $\lambda_{x,y}$ are deterministic but non-constant, as long as they satisfy uniform bounds as in assumption {\bf A.1}, i.e.\ $0< m_{\rm min} \le m_x \le m_{\rm max}$, $0\le \lambda_{x,y} \le \lambda_{\rm max}$, while disorder still enters in the form of i.i.d.\ random variables $k_x$ with assumptions as above. In fact, one can also relax the condition that the $k_x$ are identically distributed and merely needs that they are distributed with respect to densities which satisfy uniform bounds on their supports and $L^{\infty}$-norms. While we don't have explicit references for all the background results on Anderson localization in Appendix~\ref{app:Anderson} in this more general setting, it is well known and frequently discussed in the references that the existing tools allow such generalizations.

\vspace{.2cm}

(ii) Causing more technical effort, but still within the range of the existing tools to prove Anderson localization (while not considered in most of the cited references), is the case where the random variables $k_x$ have unbounded support, i.e.\ can be arbitrarily large. As a result, the operators $h_L$ do not satisfy norm bounds uniform in the disorder. In particular, $(\alpha,r)$-eigenfunction correlators as defined in (\ref{eq:dynlocgeneral}) would now also have to be considered as singular for $\alpha=1/2$, just as the $\alpha=-1/2$ eigenfunction correlators are in the non-gapped case where the $k_x$ can be arbitrarily close to $0$.

\vspace{.2cm}

(iii) More interesting would be cases where the $k_x$ are deterministic, for example a non-zero constant, while disorder enters in the form of random masses $m_x$ and/or couplings $\lambda_{x,y}$. For suitable regimes, tools such as the fractional moments method used in Appendix~\ref{app:Anderson} will likely yield localization results for the underlying single particle Hamiltonians $h_L$ with these types of random parameters. In fact, for the case of random $\lambda_{x,y}$ some results on this are provided in Section~5 of \cite{AM}.

But a major difference to the Anderson case is that these types of single particle Hamiltonians $h_L$ will have an extended ground state. In fact, if $k_x = {\rm const}$, then $h_L$ as defined by (\ref{eq:h0form}) and (\ref{onepartham}) will have the ground state $\varphi_x = Cm_x^{1/2}$. As a result, one can not expect that $h_L$ is localized near the bottom of its spectrum. It is known in the theory of one-particle random Hamiltonians  that extended states of this type, even if they only exist for a single energy level, may lead to non-trivial transport \cite{JSS}.

It is beyond the scope of our work to decide if this may lead to transport in oscillator systems, e.g.\ in
the form of lower bounds on commutators as appearing in the Lieb-Robinson bounds. But we
mention that the dispersive estimates proven in \cite{BorovykSims} for oscillator systems with
constant parameters can be interpreted as a transport property inside the light cone established by
the (non-zero velocity) Lieb-Robinson bound for such systems. We also mention that oscillator
systems with random masses have seen considerable interest in the literature for many years
\cite{Rieder1967, MatsudaIshii, CasherLebowitz}, see \cite{Ajanti2011,Bernardin2011} for some
recent references. These works are concerned with classical oscillator systems and
other types of phenomena than what we have considered here.

\section{Results in Infinite Volume} \label{sec:infinitevolume}

In this section, we discuss possible extensions of the previous results to the
infinite volume setting. For certain systems, the infinite volume single site
Hamiltonian will have a positive lower bound uniform in the randomness.
Under this assumption, the conditional results
found in Sections~\ref{sec:dlc}, \ref{sec:gsc}, and \ref{sec:tsc} can be proven directly in the infinite volume.
To see that this is true, one need only introduce the corresponding quantities of interest which,
in this special case, are obviously well-defined, see e.g.\ the discussion below. A more serious challenge is to
prove results that apply to gapless random oscillator models where the notion of
singular eigenfunction correlators is not a-priori clear. With this as a goal,
we focus solely on systems for which the single site Hamiltonian reduces to the
Anderson model. In fact, throughout this section we will assume {\bf A.3} as in
Section~\ref{sec:applications} and therefore, the results found in Appendix~\ref{app:AndersonIV}
are applicable.

We begin by introducing the CCR algebra on $\ell^2(\mathbb{Z}^d)$ and then define the
time evolution corresponding to these harmonic models as a quasi-free dynamics on this Weyl algebra.
A more thorough introduction to this general framework can be found in \cite{BratRob}.
To avoid additional complications, we focus on bounds for Weyl operators and
only comment on how to treat positions and momenta at the end.

For the Anderson model, it is proven in Appendix~\ref{app:AndersonIV}, see specifically Lemma~\ref{lem:deltaindomain},
that almost surely each $\delta_x$ is in the domain of the single site Hamiltonian.
As a result, the notion of singular eigenfunction correlators (in this gapless case) is well defined and, e.g. in the large disorder regime,
one can show that they decay appropriately. This is the main content of Appendix~\ref{app:AndersonIV}.
Consequently, a zero velocity Lieb-Robinson bound, see Theorem~\ref{thm:infvoldynloc},
follows as in Section~\ref{sec:dlc}.

The remainder of the section is used to
introduce infinite volume ground and thermal states. These states are also defined in
terms of singular objects which, in the specific gapless case where the single site Hamiltonian
is the Anderson model, we can argue they are almost surely well-defined.
Analogues of the results in Sections~\ref{sec:gsc} and \ref{sec:tsc} then follow.
We focus our attention on thermal states as the corresponding results for ground states
use similar arguments.

We start by introducing the single-particle Hamiltonian $h$ on $\ell^2( \mathbb{Z}^d)$.
Assuming {\bf A.3}, we extend the operator $h_0$, whose matrix elements are still given by
(\ref{eq:matrixelemh0}), to all of $\ell^2(\mathbb{Z}^d)$.
The single particle Hamiltonian $h$ is then defined as in (\ref{onepartham});
now on $\ell^2( \mathbb{Z}^d)$. By assumption, $h$ is self-adjoint and bounded;
it still satisfies (\ref{eq:normboundh}) uniformly in the disorder. To avoid additional prefactors,
we will also assume that the constant mass is chosen so that $\mu = I$.

For the remainder of this section, it will be convenient to identify
$\ell^2(\mathbb{Z}^d)= \ell^2(\mathbb{Z}^d; \mathbb{C})$ with
$\ell^2(\mathbb{Z}^d; \mathbb{R}) \oplus \ell^2(\mathbb{Z}^d; \mathbb{R})$ and regard
\begin{equation}
f \in \ell^2( \mathbb{Z}^d; \mathbb{C}) \quad \mbox{as} \quad f = \left( \begin{array}{c} {\rm Re}[f] \\ {\rm Im}[f] \end{array} \right) \in \ell^2(\mathbb{Z}^d; \mathbb{R}) \oplus \ell^2(\mathbb{Z}^d; \mathbb{R}) \, .
\end{equation}
With this identification, we can more easily express dependencies on real and imaginary parts
in terms of matrix multiplication, see below.

We now recall the definition of the Weyl algebra over $\ell^2( \mathbb{Z}^d)$.
To each $f \in \ell^2(\mathbb{Z}^d)$, associate a non-zero Weyl operator
$W(f)$ satisfying
\begin{equation}
W(f)^* = W(-f) \quad \mbox{for each } f \in \ell^2(\mathbb{Z}^d),
\end{equation}
and
\begin{equation}
W(f) W(g) = e^{- i \sigma(f,g) /2}W(f+g)
\quad \mbox{for all } f, g \in \ell^2(\mathbb{Z}^d) \, ,
\end{equation}
where
\begin{equation}
\sigma(f,g) = {\rm Im}[ \langle f,g \rangle] = (J f, g ),  \quad J = \left( \begin{array}{cc} 0 & - I \\ I & 0 \end{array} \right)\, ,
\end{equation}
and $\langle \cdot , \cdot \rangle$, resp.\ $( \cdot, \cdot)$, denotes the inner product on the complex (resp.\ real) $\ell^2$-space.
It is well-known, see e.g.\ Theorem 5.2.8 of \cite{BratRob}, that there is a unique, up to $*$-isomorphism, $C^*$-algebra
generated by the Weyl operators with the property that $W(0) = I$, $W(f)$ is unitary for all $f \in \ell^2(\mathbb{Z}^d)$, and
$\| W(f) - I \| = 2$ for all $f \in \ell^2(\mathbb{Z}^d) \setminus \{0 \}$. This algebra is called the CCR algebra, or the Weyl algebra, over
$\ell^2(\mathbb{Z}^d)$ and we will denote it by $\mathcal{W}$.

A quasi-free dynamics on $\mathcal{W}$, see e.g.\ Theorem 5.2.8 (4) \cite{BratRob}, is a one-parameter group of $*$-automorphisms $\tau_t$
of the form
\begin{equation} \label{eq:quasi}
\tau_t(W(f)) = W(T_tf) \quad \mbox{for all } f \in \ell^2( \mathbb{Z}^d)
\end{equation}
where $T_t: \ell^2(\mathbb{Z}^d) \to \ell^2(\mathbb{Z}^d)$ is a group of real-linear, symplectic transformations, i.e.
\begin{equation} \label{eq:impartpres}
\left(JT_t f, T_t g \right)  = \left( f, g \right)  \quad \mbox{for all } f, g \in \ell^2(\mathbb{Z}^d) \quad \mbox{and } t \in \mathbb{R}.
\end{equation}

Motivated by the finite-volume expressions (\ref{eq:reft}) and (\ref{eq:imft}) in Section~\ref{sec:dlc},
we define this infinite volume harmonic dynamics on $\mathcal{W}$, as in (\ref{eq:quasi}) above,
in terms of the mapping $T_t$ given by
\begin{equation} \label{eq:defTtiv}
T_t = \left(  \begin{array}{cc} \cos(2t h^{1/2}) & - h^{1/2} \sin( 2t h^{1/2}) \\  h^{-1/2} \sin(2t h^{1/2})  &  \cos(2th^{1/2})  \end{array} \right) \,
\end{equation}
where $h$ is the infinite volume single-particle Hamiltonian previously discussed, and we recall that we have chosen constants so that
$\mu = I$. Since all entries in (\ref{eq:defTtiv}) are bounded functions of $h$, $T_t$ defines a
bounded, real-linear transformation on $\ell^2(\mathbb{Z}^d)$ for each $t \in \mathbb{R}$ by the usual functional
calculus for self-adjoint operators. Moreover, it is easy to see that $T_t$ is a
group and,  one readily checks that
\begin{equation}
T_t^T J T_t = J\, ,
\end{equation}
i.e.\ for each $t \in \mathbb{R}$, $T_t$ is symplectic.
As is indicated above, this is sufficient to define the harmonic evolution as a quasi-free
dynamics on $\mathcal{W}$.

We can now prove the analogue of Theorem~\ref{thm:sdlweylcom} in this setting.
Recall that in the notation of {\bf A.3}, $\rho$ is the density of the random variables and that large disorder is expressed in terms of smallness of $\|\rho\|_{\infty}$.

\begin{theorem} \label{thm:infvoldynloc}
Let $\| \rho \|_{\infty}$ be sufficiently small. Then there exist $C<\infty$ and $\mu>0$ such that
\begin{equation} \label{eq:infvoldynloc}
\E \left( \sup_{t\in \R} \| [ \tau_t(W(f)), W(g)]\| \right) \le C \sum_{x,y} |f(x)||g(y)| e^{-\mu |x-y|}
\end{equation}
for all $f, g \in \ell^2(\Z^d)$.
\end{theorem}

\begin{proof}
As is discussed in the beginning of Appendix~\ref{app:AndersonIV}, an infinite volume
version of Proposition~\ref{prop:FML} is well-known to hold. Thus for any $d \geq1$ and $\| \rho \|_{\infty}$ sufficiently
small, Proposition~\ref{prop:FML}(b) holds and so the assumption of Proposition~\ref{prop:EFCboundIVbc}(a)
is satisfied on an interval containing the (uniformly bounded) spectrum of $h$.
In this case, the proof of (\ref{eq:infvoldynloc}) immediately follows using the arguments of Theorem~\ref{thm:sdlweylcom}
with (\ref{eq:EFCboundIV2}) as the new input.
\end{proof}

We now turn our attention to states and correlation bounds.
To begin, we review how these infinite volume states are rigorously introduced.
We first sketch the details for thermal states. The corresponding results for ground states immediately follow.
Again, more general information, including these facts, can be found e.g. in \cite{BratRob}.

Fix $\beta >0$ and consider the functional on $\mathcal{W}$ defined by setting
\begin{equation}
\omega_{\beta}(W(f)) = e^{- \frac{1}{2} s_{\beta}(f,f)} \quad \mbox{for all } f \in \ell^2(\mathbb{Z}^d)
\end{equation}
where $s_{\beta}$ is the quadratic form given by
\begin{equation}
s_{\beta}(f,f) = \left\{ \begin{array}{cc} \frac{1}{2} \| M_{\beta} f \|^2 & \mbox{for all } f \in \mathcal{D}_{\beta} \\
+ \infty & \mbox{otherwise} \end{array} \right.
\end{equation}
with
\begin{equation} \label{eq:defMb}
M_{\beta} =  \left( \begin{array}{cc} h^{-1/4} \coth(\beta h^{1/2})^{1/2} & 0 \\ 0 & h^{1/4} \coth(\beta h^{1/2})^{1/2} \end{array} \right)
\end{equation}
and form domain
\begin{equation}
\mathcal{D}_{\beta} = \left\{ f \in \ell^2(\mathbb{Z}^d) \, : \, {\rm Re}[f] \in D \left( h^{-1/2} \right) \right\} \, ,
\end{equation}
where $D(h^{-1/2})$ is the domain of $h^{-1/2}$, see e.g.\ Lemma~\ref{lem:hinv} and the preceding discussion.
Note that if $f \in \mathcal{D}_{\beta}$, then $h^{-1/4} \coth(\beta h^{1/2})^{1/2}{\rm Re}[f]$ is well-defined since
$h^{1/4}  \coth(\beta h^{1/2})^{1/2}$ is bounded and all these functions of $h$ commute. Since $h$ is random, so too
are $\omega_{\beta}$ and its domain $\mathcal{D}_{\beta}$, however, given Lemma~\ref{lem:deltaindomain},
we know that almost surely $\mathcal{F}$, the (deterministic) set of all functions in $\ell^2(\mathbb{Z}^d)$ with finite support, is
in $\mathcal{D}_{\beta}$.

To see that the functional $\omega_{\beta}$ actually defines a
state on $\mathcal{W}$, i.e. a positive linear functional, we need only check that
\begin{equation}
\sigma(f,g)^2 \leq 4 s_{\beta}(f,f) s_{\beta}(g,g)
\end{equation}
see e.g. \cite{Ver}. This follows by noting that the matrix
\begin{equation}
M = \left( \begin{array}{cc} h^{-1/4} & 0 \\ 0 & h^{1/4} \end{array} \right)
\end{equation}
is symplectic, that $\| M f \| \leq \|M_{\beta}f \|$ for all $f \in \mathcal{D}_{\beta}$, and therefore
\begin{equation}
\sigma(f,g)^2 = \sigma(Mf, Mg)^2 \leq \|M f \|^2 \|M g \|^2 \leq 4 s_{\beta}(f,f) s_{\beta}(g,g) \, .
\end{equation}

We next note that $\omega_{\beta}$ is time invariant, i.e., the fact that
\begin{equation} \label{eq:timeinvIV}
\omega_{\beta}(\tau_t(W(f))) = \omega_{\beta}(W(f)) \quad \mbox{for all } f \in \ell^2(\mathbb{Z}^d) \, .
\end{equation}
First, observe that if $f \in \mathcal{D}_{\beta}$, then $T_tf \in \mathcal{D}_{\beta}$ for all $t \in \mathbb{R}$.
This follows by inspection of the mapping $T_t$, see (\ref{eq:defTtiv}). Since $T_t$ is invertible, this
verifies (\ref{eq:timeinvIV}) for all $f \in \ell^2(\mathbb{Z}^d) \setminus \mathcal{D}_{\beta}$.
Otherwise, a short calculation shows that
\begin{equation}
M_{\beta}T_t = R_t M_{\beta} \quad \mbox{where} \quad R_t = \left( \begin{array}{cc} \cos(2t h^{1/2}) & - \sin(2t h^{1/2}) \\ \sin(2t h^{1/2}) & \cos(2t h^{1/2}) \end{array} \right)
\end{equation}
which is clearly orthogonal for all $t \in \mathbb{R}$. (\ref{eq:timeinvIV}) is proven.

We can now state our results on dynamic correlations for the infinite volume thermal states.
\begin{theorem} \label{thm:infvoldyntscor}
Let $\| \rho \|_{\infty}$ be sufficiently small. Then there exist $C<\infty$ and $\mu>0$ such that
\begin{equation} \label{eq:infvoldyntscor}
\begin{split}
\E \left( \sup_{t\in \R} \left| \omega_{\beta} \left( \tau_t(W(f)) W(g)\right)  - \omega_{\beta} \left( \tau_t(W(f))\right)\omega_{\beta} \left( W(g)\right)\right| \right) \\
\quad \le C \sum_{x,y} |f(x)|^{1/2} |g(y)|^{1/2} e^{-\mu |x-y|}
\end{split}
\end{equation}
for all $f, g \in \mathcal{F}$.
\end{theorem}

\begin{proof}
As in the previous theorem, it is clear that for $\| \rho \|_{\infty}$ sufficiently small, the
assumptions of Proposition~\ref{prop:verysingEFCIV}(a) hold on an interval containing the spectrum of $h$.
In this case, one sees that the argument for the proof of Theorem~\ref{thm:thermalweylcor} (with $r=1/2$) carries through,
using the results of the above discussion and (\ref{eq:SEFCbound1IV}) as input, to show (\ref{eq:infvoldyntscor}).
\end{proof}

In addition, we also have the follow result on static correlations, which does not
require the large-disorder assumption but only depends on localization near the bottom
of the spectrum of the single-particle Hamiltonian.

\begin{theorem} \label{thm:infvolstattscor}
There exist $C<\infty$ and $\mu>0$ such that
\begin{equation} \label{eq:infvoltscor2}
\E \left( \left| \omega_{\beta} \left( W(f) W(g)\right)  - \omega_{\beta} \left(W(f)\right)\omega_{\beta} \left( W(g)\right)\right| \right)
 \le C \sum_{x,y} |f(x)|^{1/2} |g(y)|^{1/2} e^{-\mu |x-y|}
\end{equation}
holds for all $f,g \in \mathcal{F}$ with disjoint support.
\end{theorem}

\begin{proof}
One easily checks that an analogue of Corollary~\ref{cor:staticthermalWeylcor} holds.
Now, for these static bounds, we use Proposition~\ref{prop:verysingEFCIV}(b) with $\varphi_1(h) = \varphi_2(h) = h^{-1/2}$.
It is now clear that (\ref{eq:infvoltscor2}) follows from the analogue of (\ref{eq:staticthermalWeylcor}) using (\ref{eq:EFCbound4IV}).
\end{proof}

Similar results, i.e. analogues of Theorem~\ref{thm:infvoldyntscor} and Theorem~\ref{thm:infvolstattscor}, hold for ground states.
Note that ground states can be defined as the $\beta \to \infty$ limit of thermal states, or, more concretely, by setting $\coth( \beta h^{1/2}) = 1$
in (\ref{eq:defMb}). For dynamic and static correlation estimates in this less singular state, we use
Proposition~\ref{prop:EFCboundIVbc}(a) and (b) respectively. We leave the details to the reader.

The ground state and thermal states of the infinite systems considered here are sufficiently
regular (see \cite[pp 37--38]{BratRob}), so that one can define position and momentum operators
in the GNS representation almost surely. As a consequence,  one can verify that, on a dense
subset of the GNS Hilbert space, formulae identical to those in
Lemma~\ref{lem:commat} hold. On this subset,  an analogue of Theorem~\ref{thm:sdlp+qcom}
can be proven, again under the large-disorder assumption.
Analogues of the results for correlations of position and momentum operators
in the ground state and thermal state similarly hold in the GNS representations.

\section{Conclusion} \label{sec:conclusion}

We conclude with a brief summary of what we consider to be the most important novel contributions of our work.

We prove dynamical localization in the form of a zero-velocity Lieb-Robinson bound for harmonic oscillator systems. Disorder provides the crucial mechanism for this, as it is necessary that the associated one-particle Hamiltonian is localized at all energies.

We prove exponential decay for ground state and thermal state correlations of disordered oscillator systems. For this it is not required that the oscillator system is gapped (as in related previous results for deterministic systems). Instead we exploit a ``mobility gap'' expressed in terms of localization of the one-particle Hamiltonian near the bottom of its spectrum.

In cases where the mobility gap extends to all energies, as for the dynamical localization bounds, we can go beyond static correlations and get exponential clustering for dynamic ground and thermal state correlations.

The extensions of our results to oscillator systems over the full Euclidean lattice provide localization results for oscillator systems in infinite volume.

In order to cover non-gapped as well as infinite volume oscillator systems, we need new results on Anderson localization. Specifically, we prove exponential decay of some types of {\it singular} eigenfunction correlators for the Anderson model required for our applications.

\appendix

%
%
%

\section{Some Results on Anderson Localization} \label{app:Anderson}

We consider the finite-volume Anderson model on $\ell^2(\Lambda_L)$, $\Lambda_L = [-L,L]^d \cap \Z^d$,
\begin{equation} \label{eq:Anderson}
h_L = h_{0,L} + V_{\omega}.
\end{equation}
Here $h_{0,L}$ is the graph Laplacian on $\Lambda_L$, given in terms of its quadratic form by
\begin{equation} \label{eq:GraphLap}
\langle f, h_{0,L} g \rangle = \sum_{\{x,y\} \subset \Lambda_L \atop |x-y|=1} \overline{(f(y)-f(x))} (g(y)-g(x)),
\end{equation}
and
\begin{equation} \label{eq:ranpot}
(V_{\omega}f)(x) = \omega_x f(x)
\end{equation}
for an array $(\omega_x)_{x\in \Z^d}$ of i.i.d.\ random variables. We assume that their common distribution is absolutely continuous with bounded and compactly supported density $\rho$. We will assume $\min(\mbox{supp}\,\rho)=0$. Thus for any given $x$ the event $\omega_x=0$ has probability zero. This implies that almost surely $h_L> 0$ (i.e.\ does not have eigenvalue $0$) and thus is invertible. Nevertheless, the bottom of the almost sure spectrum of the infinite volume Anderson model $h_0+V_{\omega}$ ($h_0$ the graph Laplacian on $\Z^d$) is $0$, see \cite{Stolz11} for an elementary proof.

This latter choice of normalization as well as the choice of the graph Laplacian (\ref{eq:GraphLap}) as background operator are motivated by our applications in earlier sections. Other background operators could be used for all the localization results stated below without affecting the proofs. In particular, as done in most of the references given below, one could work with the restriction $\chi_{\Lambda_L} h_0 \chi_{\Lambda_L}$ of the infinite volume Laplacian to $\Lambda_L$, which differs from $h_{0,L}$ by a boundary condition ($h_{0,L}$ is sometimes called the discrete Neumann Laplacian) and an energy shift (if $h_0$ is chosen as the next-neighbor-hopping or adjacency operator). We could also use periodic boundary conditions in (\ref{eq:GraphLap}), i.e.\ replace $\Lambda_L$ by a $d$-dimensional discrete torus. In the latter case distances of lattice sites in (\ref{eq:GraphLap}) as well as in the results below have to be interpreted as distances on the torus.

The main reason for requiring that the distribution of the $\omega_x$ has bounded and compactly supported density $\rho$ is that this allows us to refer to the rather strong localization properties of the Anderson model which have been proven by the fractional moment method (FMM), see \cite{AM, Aizenman94, Graf, ASFH} and, for a recent survey, \cite{Stolz11}. Some of the results proven by this method hold under the weaker assumption of H\"older continuous distributions, e.g.\ \cite{ASFH}, and the assumption of compact support of $\rho$ could also be relaxed. But the method can not be extended to more singular distributions and other methods do not yield equally strong results.

The main technical characteristic of the FMM is that it initially aims at proving {\it fractional moment localization}, from which other properties such as spectral and dynamical localization are derived as consequences.

For given $0<s<1$ and an interval $I \subset \R$, we say that the finite-volume Anderson model $h_L$ has localized $s$-moments in $I$ if there are constants $C'<\infty$ and $\mu>0$ such that
\begin{equation} \label{eq:FML}
\E \left( |G_L(x,y;E+i\varepsilon)|^s \right) \le C' e^{-\mu |x-y|}
\end{equation}
for all positive integers $L$, all $x,y \in \Lambda_L$, $E \in I$ and $\varepsilon \in \R$. Here $G_L(x,y;z) := \langle \delta_x, (h_L-z)^{-1} \delta_y \rangle$ is the Green function for $h_L$ and $\E(\cdot)$ refers to the disorder average.

The following summarizes the regimes in which fractional moment localization for the finite volume Anderson model has been shown.

\begin{proposition} \label{prop:FML}
The $d$-dimensional finite volume Anderson model $h_L$ has localized $s$-moments in $I$ under each of the following assumptions:

(a) $d\ge 1$ and $s\in(0,1)$ arbitrary, and $I=[0,E_0]$ for some $E_0 = E_0(s,d)>0$,

(b) $d\ge 1$ and $s\in(0,1)$ arbitrary, $\|\rho\|_{\infty} \le \rho_{max}$ for $\rho_{max} = \rho_{max}(s,d)>0$ sufficiently small, and $I=[0,\infty)$,

(c) $d=1$, $s>0$ sufficiently small, and $I=[0,\infty)$.
\end{proposition}

Historically, the first localization result obtained by using the FMM was for the large disorder regime (b), see \cite{AM} and \cite{Graf}. In particular, the statement in (b) covers the case of an Anderson model $h_{0,L} +\lambda V_{\omega}$, where $V_{\omega}$ is given via a fixed density $\rho$ and the coupling parameter $\lambda$ is sufficiently large. Band edge fractional moment localization (a) was obtained in \cite{ASFH}. That (c) follows from the well known properties of the Lyapunov exponents of the one-dimensional Anderson model was first observed in \cite{Minami96}, see also \cite{HSS10}. Much of this, including the proofs, is reviewed in \cite{Stolz11}.

One of the virtues of working in finite volume is that $\varepsilon=0$ is allowed in (\ref{eq:FML}). The proofs show that any fixed $E\in \R$ is an eigenvalue of $h_L$ with probability zero and thus, as the spectrum of $h_L$ is discrete, $(h_L-E)^{-1}$ exists almost surely and the left hand side of (\ref{eq:FML}) makes sense for $\varepsilon=0$.

Among the consequences of (\ref{eq:FML}), the one of most interest to us here is the following: We say that the Anderson model $h_L$ has localized eigenfunction correlators in an interval $I \subset \R$, if there exist $\mu>0$ and $C<\infty$ such that
\begin{equation} \label{eq:AndEFC}
\E \left( \sup_{|u|\le 1} | \langle \delta_x, u(h_L) \chi_I(h_L) \delta_y \rangle| \right) \le C e^{-\mu |x-y|}
\end{equation}
for all positive integers $L$ and $x,y \in \Lambda_L$. The supremum is taken over all functions $u:\R \to \C$ which satisfy the pointwise bound $|u|\le 1$ on $I$. The operators $u(h_L)$ and $\chi_I(h_L)$ are defined via the functional calculus for self-adjoint operators, in particular $\chi_I(h_L)$ is the spectral projection for $h_L$ onto the interval $I$. Of course, as we work in finite volume, the functional calculus comes down to simple finite eigenfunction expansions. For this reason we also don't need any measurability assumptions for the functions $u$.

For more explanation of the reason for describing the left hand side of (\ref{eq:AndEFC}) as an eigenfunction correlator see \cite{Aizenman94, ASFH, Stolz11}. The choice $u_t(E) = e^{-itE}$, $t\in \R$, shows that (\ref{eq:AndEFC}) implies the dynamical localization bound
\begin{equation} \label{eq:AndDL}
\E \left( \sup_{t\in \R} | \langle \delta_x, e^{-ith_L} \chi_I(h_L) \delta_y \rangle| \right) \le C e^{-\mu |x-y|}.
\end{equation}
While this explains why (\ref{eq:AndEFC}) is sometimes referred to as dynamical localization, we will need (\ref{eq:AndEFC}) also for other types of functions $u$.

The crucial bound which allows to deduce localization of eigenfunction correlators from fractional moment localization is the content of the next result.

\begin{proposition} \label{prop:EFCFMbound}
For every $0<s<1$, every dimension $d$, and every bounded and compactly supported probability density $\rho$, there exists $C'' = C''(s,d,\rho)<\infty$ such that
\begin{equation} \label{eq:EFCFMbound}
\E \left( \sup_{|u|\le 1} | \langle \delta_x, u(h_L) \chi_I(h_L) \delta_y \rangle | \right) \le C'' \left( \int_I \E(|G_L(x,y;E)|^s)\,dE \right)^{\frac{1}{2-s}}
\end{equation}
for every interval $I\subset \R$.
\end{proposition}

This relation originates from the works \cite{Aizenman94, ASFH}, a detailed proof can also be found in \cite{Stolz11}. By combining Propositions~\ref{prop:FML} and \ref{prop:EFCFMbound} one gets the first part of the following result.

\begin{proposition} \label{prop:EFCbounds}
Let $h_L$ be the finite-volume Anderson model defined by (\ref{eq:Anderson}), (\ref{eq:GraphLap}) and (\ref{eq:ranpot}) with assumptions as above.

(a) If $h_L$ has localized $s$-moments in $I$ in the sense of (\ref{eq:FML}), then
\begin{equation} \label{eq:EFCbound1}
\E \left( \sup_{|u|\le 1} | \langle \delta_x, u(h_L) \chi_J(h_L) \delta_y \rangle| \right) \le C |J|^{1/(2-s)} e^{-\mu|x-y|/(2-s)}
\end{equation}
for every subinterval $J$ of $I$, every positive integer $L$ and all $x,y \in \Lambda_L$. Here $C = C'' (C')^{1/(2-s)}$ with $C'$ from (\ref{eq:FML}) and $C''$ from (\ref{eq:EFCFMbound}).

(b) If $h_L$ is $s$-localized in $[0,E_0]$ in the sense of (\ref{eq:FML}) and $\alpha > -s/(2-s)$, then there exists $C_1<\infty$ such that
\begin{equation} \label{eq:EFCbound2}
\E \left( \sup_{|u|\le 1} | \langle \delta_x, h_L^{\alpha} u(h_L) \chi_{[0,E_0]}(h_L) \delta_y \rangle | \right) \le C_1 e^{-\mu|x-y|/(2-s)}
\end{equation}
for all $L$ and $x,y \in \Lambda_L$.

(c) Suppose that $\varphi:(0,\infty) \to \C$ satisfies $|\varphi(t)| \le Ct^{\alpha}$ for $t$ near $0$ with some $C<\infty$ and $\alpha>-1$, and that $\varphi$ has an analytic extension to a semi-strip $\{z: \mbox{Re}\,z >0, |\mbox{Im}\,z|<\eta\}$ for some $\eta>0$. Then there exist $C_2<\infty$ and $\mu_2>0$ such that
\begin{equation} \label{eq:EFCbound3}
\E \left( |\langle \delta_x, \varphi(h_L) \delta_y \rangle | \right) \le C_2 e^{-\mu_2 |x-y|}
\end{equation}
for all $L$ and $x,y \in \Lambda_L$.

\end{proposition}

The statements in (b) and (c) involve the terms $h_L^{\alpha}$ and $\varphi(h_L)$ which are allowed to be singular at energy $E=0$. Also, the left hand side of (c) does not require to include the projection $\chi_{[0,E_0]}(h_L)$ onto the localized energy regime as in (b). The price for this is that one can not include a supremum over a class of functions of $h_L$ as in (a) and (b) and, in particular, say nothing about dynamics.

Below we prove (b) using (a) and a Riemann sum argument. The proof of (c) is done by contour integration, similar to methods previously used in \cite{AizenmanGraf}.

\begin{proof}[Proof of Propositon~\ref{prop:EFCbounds}] It remains to prove (b) and (c).

If $\alpha\ge 0$, then (\ref{eq:EFCbound2}) follows easily from (\ref{eq:EFCbound1}) with $J=[0,E_0]$, as in this case $x^{\alpha} \le E_0^{\alpha}$ on $[0,E_0]$.

For $-s/(2-s)<\alpha <0$, decompose the interval $(0,E_0]$ into $I_n := (E_0/(n+1), E_0/n]$, $n=1,2,\ldots$. As $0$ is almost surely not an eigenvalue of $h_L$, we have with probability one that
\begin{equation} \label{eq:crucialsplit}
\chi_{[0,E_0]}(h_L) = \chi_{(0,E_0]}(h_L) = \sum_{n=1}^{\infty} \chi_{I_n}(h_L).
\end{equation}
Convergence is trivial here as, due to the discreteness of the spectrum of $h_L$, the sum is finite.
If $|u|\le 1$, then $|x^{\alpha} u(x)| \le (E_0/(n+1))^{\alpha}$ on $I_n$. Thus it follows from (a) that
\begin{equation} \label{eq:sumbound}
\E \left( \sup_{|u|\le 1} | \langle \delta_x, h_L^{\alpha} u(h_L) \chi_{[0,E_0]}(h_L) \delta_y \rangle| \right) \le \sum_{n=1}^{\infty} C \left( \frac{E_0}{n+1} \right)^{\alpha} |I_n|^{1/(2-s)} e^{-\mu |x-y|/(2-s)}.
\end{equation}
As $\left( \frac{E_0}{n+1}\right)^{\alpha} |I_n|^{1/(2-s)} \sim \left( \frac{1}{n} \right)^{\alpha+\frac{2}{2-s}}$ and $\alpha + \frac{2}{2-s}>1$, the series converges.  We get (\ref{eq:EFCbound2}) with $C_1 = C \sum_{n=1}^{\infty} \left( \frac{E_0}{n+1} \right)^{\alpha} |I_n|^{1/(2-s)}$.

To prove (c), we first choose $s<1$ such that $\alpha > -s/(2-s)$ and then $E_0>0$ as in Proposition~\ref{prop:FML}(a). Decompose
\begin{eqnarray} \label{eq:partcdecomp}
\E \left( |\langle \delta_x, \varphi(h_L) \delta_y \rangle | \right) & \le & \E \left( | \langle \delta_x, \varphi(h_L) \chi_{[0,E_0]}(h_L) \delta_y \rangle| \right) \\
& & \mbox{} + \E \left( | \langle \delta_x, \varphi(h_L) \chi_{(E_0,\infty)}(h_L) \delta_y \rangle | \right). \nonumber
\end{eqnarray}

For the first term we get the required bound from (b), using the bound $|\varphi(t)| \le Ct^{\alpha}$. For the second term we argue as follows:

As $\rho$ is compactly supported there exists $M<\infty$ such that $h_L \le M$ uniformly in $L$ and the disorder. Assume that $E_0$ is not an eigenvalue of $h_L$ (which holds almost surely). Then, with $E(\cdot)$ denoting the spectral resolution of $h_L$,
\begin{eqnarray} \label{eq:Cauchyint}
\langle \delta_x, \varphi(h_L) \chi_{(E_0,\infty)}(h_L) \delta_y \rangle & = & \int_{(E_0,M]} \varphi(t) \,d \langle \delta_x, E(t) \delta_y \rangle \\
& = & \int_{(E_0,M]} \frac{1}{2\pi i} \int_{\Gamma} \frac{\varphi(z)}{z-t} \,dz \, d\langle \delta_x, E(t) \delta_y \rangle \nonumber
\end{eqnarray}
by the Cauchy integral formula, where we have used that $\varphi$ has an analytic extension to a semi-strip and thus we can choose for $\Gamma$ the rectangular contour with vertices $E_0-i\eta/2$, $M+1-i\eta/2$, $M+1+i\eta/2$ and $E_0+i\eta/2$.

We work in finite volume, so that the spectral integral in (\ref{eq:Cauchyint}) is a finite sum, allowing for a trivial exchange of integration order. One gets
\begin{eqnarray} \label{eq:Greenbound}
| \langle \delta_x, \varphi(h_L) \chi_{(E_0,\infty)}(h_L) \delta_y \rangle | & = & \left| -\frac{1}{2\pi i} \int_{\Gamma} \varphi(z) \langle \delta_x, (h_L-z)^{-1} \chi_{(E_0,M]}(h_L) \delta_y \rangle \,dz \right| \\
& \le & C \int_{\Gamma} |\langle \delta_x, (h_L-z)^{-1} \chi_{(E_0,M]}(h_L) \delta_y \rangle |\,|dz|, \nonumber
\end{eqnarray}
with $C =\frac{1}{2\pi} \max \{ |\varphi(z)|: z\in \Gamma\}$.

Decompose $\Gamma = \Gamma_1 \cup \Gamma_2$, where $\Gamma_1$ is the line segment from $E_0+i\eta/2$ to $E_0-i\eta/2$ and $\Gamma_2$ the remaining part of the contour. Choose $0<s'<1/2$ and use the trivial bound $\|(h_L-(E_0+ia))^{-1}\| \le 1/a$ to estimate
\begin{eqnarray} \label{eq:morecalc} \\
\lefteqn{\E \int_{\Gamma_1} | \langle \delta_x, (h_L-z)^{-1} \chi_{(E_0,M]}(h_L) \delta_y \rangle |\,|dz|} \nonumber \\ & \le & \int_{-\eta/2}^{\eta/2} a^{-(1-s')} \E \left( | \langle \delta_x, (h_L-(E_0+ia))^{-1} \chi_{(E_0,M]}(h_L) \delta_y \rangle|^{s'} \right)\,da \nonumber \\
& \le & \int_{-\eta/2}^{\eta/2} a^{-(1-{s'})} \E \left( \sum_v |G_L(x,v;E_0+ia)|^{s'} |\langle \delta_v, \chi_{(E_0,M]}(h_L) \delta_y \rangle |^{s'} \right) \,da \nonumber \\
& \le & \int_{-\eta/2}^{\eta/2} a^{-(1-{s'})} \sum_v \left\{ (\E |G_L(x,v;E_0+ia)|^{2{s'}})^{1/2} (\E|\langle \delta_v, \chi_{(E_0,M]}(h_L) \delta_y \rangle |^{2{s'}})^{1/2} \right\}\,da. \nonumber
\end{eqnarray}

By Proposition~\ref{prop:FML}, $\E(|G_L(x,v;E_0+ia)|^{2{s'}}) \le C_1 e^{-\mu_1|x-v|}$, while
\begin{eqnarray} \label{eq:morecalc2}
\E(|\langle \delta_v, \chi_{(E_0,M]}(h_L) \delta_y \rangle |^{2{s'}}) & \le & \E (|\langle \delta_v, \chi_{(E_0,M]}(h_L) \delta_y \rangle |) \\
& = & \E (|\langle \delta_v, (I-\chi_{[0,E_0]}(h_L)) \delta_y \rangle | \nonumber \\
& \le & \delta_{vy} + C_2 e^{-\mu_2|v-y|} \le (C_2+1) e^{-\mu_2 |v-y|} \nonumber
\end{eqnarray}
by (a). We can thus bound the right hand side of (\ref{eq:morecalc}) by
\begin{equation} \label{eq:morecalc3}
C\int_{-\eta/2}^{\eta/2} a^{-(1-{s'})} \sum_v e^{-\mu_1|x-v|/2} e^{-\mu_2|v-y|/2} \le C' e^{-\mu |x-y|}
\end{equation}
for any $\mu < \frac{1}{4} \min\{\mu_1, \mu_2\}$. A similar bound is found for the second term contributing to (\ref{eq:Greenbound}), corresponding to $\Gamma_2$, where one can use the well-known deterministic Combes-Thomas bound $|G_L(x,v;z)| \le C_1 e^{-\mu_1|x-v|}$, which holds uniformly in $z\in \Gamma_2$. For a self-contained proof of the latter, which extends to the finite volume case considered here, see, e.g., Section~11.2 of \cite{Kirsch}.

All of this can now be combined to get the desired decay bound for the second term in (\ref{eq:partcdecomp}).

\end{proof}

Parts (b) and (c) of Proposition~\ref{prop:EFCbounds} exclude the case $\alpha =-1$. As our final result in this appendix we show how these stronger zero-energy singularities can be handled by including a fractional moment (the $1/2$ moment) in the expectation of the eigenfunction correlator.

\begin{proposition} \label{prop:verysingEFC} Fix assumptions as above.

(a) If $h_L$ has $s$-localized moments in $[0,E_0]$ and $\alpha > -1- \frac{s}{2-s}$, then there exist $C_3<\infty$ and $\mu_3>0$ such that
\begin{equation} \label{eq:SEFCbound1}
\E \left( \sup_{|u|\le 1} | \langle \delta_x, h_L^{\alpha} u(h_L) \chi_{[0,E_0]}(h_L) \delta_y \rangle |^{1/2} \right) \le C_3 e^{-\mu_3 |x-y|}
\end{equation}
for all $L$ and $x,y \in \Lambda_L$.

(b) Suppose that $\varphi_1$ and $\varphi_2$ both have the properties of the function $\varphi$ in Proposition~\ref{prop:EFCbounds}. Then there exist $C_4<\infty$ and $\mu_4>0$ such that
\begin{equation} \label{eq:EFCbound4}
\E \left( |\langle \delta_x, \varphi_1(h_L) \varphi_2(h_L) \delta_y \rangle |^{1/2} \right) \le C_4 e^{-\mu_4 |x-y|}
\end{equation}
for all $L$ and $x,y \in \Lambda_L$.

\end{proposition}

\begin{proof} Choose $\alpha_1 >-1$ and $\alpha_2 > -s/(2-s)$ such that $\alpha = \alpha_1 + \alpha_2$. Then
\begin{eqnarray}
\langle \delta_x, h_L^{\alpha} u(h_L) \chi_{[0,E_0]}(h_L) \delta_y \rangle & = & \langle h_L^{\alpha_1} \delta_x, h_L^{\alpha_2} u(h_L) \chi_{[0,E_0]}(h_L) \delta_y \rangle \\
& = & \sum_w \langle h_L^{\alpha_1} \delta_x, \delta_w \rangle \langle \delta_w, h_L^{\alpha_2} u(h_L) \chi_{[0,E_0]}(h_L) \delta_y \rangle. \nonumber
\end{eqnarray}

Using Cauchy-Schwarz on the expectation we find from this that
\begin{eqnarray}
\lefteqn{\E \left( \sup_{|u|\le 1} | \langle \delta_x, h_L^{\alpha} u(h_L) \chi_{[0,E_0]}(h_L) \delta_y \rangle|^{1/2} \right)} \\
& \le & \sum_w \left( \E(|\langle \delta_x, h_L^{\alpha_1} \delta_w \rangle|) \right)^{1/2} \left( \E( \sup_{|u|\le 1} |\langle \delta_w, h_L^{\alpha_2} u(h_L) \chi_{[0,E_0]}(h_L) \delta_y \rangle|) \right)^{1/2}. \nonumber
\end{eqnarray}
Part (c) of Proposition~\ref{prop:EFCbounds} applies to the first term of the sum, while part (b) applies to the second term, giving exponentially decaying bounds in $|x-w|$ and $|w-y|$, respectively, which carry over with a suitably reduced rate to the convolution-type sum. This proves (a).

For (b) one argues similarly, bounding
\begin{eqnarray}
\lefteqn{ \E \left( | \langle \delta_x, \varphi_1(h_L) \varphi_2(h_L) \delta_y \rangle |^{1/2} \right)} \\
& \le & \sum_w \left( \E (|\langle \delta_x, \varphi_1(h_L) \delta_w \rangle|) \right)^{1/2} \left( \E(|\langle \delta_w, \varphi_2(h_L) \delta_y \rangle|) \right)^{1/2}, \nonumber
\end{eqnarray}
and using Proposition~\ref{prop:EFCbounds}(c) on both terms to conclude as above.

\end{proof}

\section{The Anderson model in infinite volume} \label{app:AndersonIV}

Our goal here is to extend the results of Appendix~\ref{app:Anderson} to the Anderson model in infinite volume. On $\ell^2(\Z^d)$ consider
\begin{equation} \label{eq:AndersonIV}
h = h_0 + V_{\omega}.
\end{equation}
The graph Laplacian $h_0$ on $\Z^d$ is given by
\begin{equation} \label{eq:GraphLap1}
\langle f, h_0 g \rangle = \sum_{\{x,y\} \subset \Z^d \atop |x-y|=1} \overline{(f(y)-f(x))} (g(y)-g(x)),
\end{equation}
and the random potential is defined by (\ref{eq:ranpot}) with i.i.d.\ random variables $(\omega_x)_{x\in \Z^d}$. As before we assume that they are distributed according to a bounded and compactly supported density $\rho$ with $\min(\mbox{\rm supp}\,\rho) =\omega_{\rm min} \ge 0$. By well known facts (e.g.\ \cite{CFKS}) the spectrum of $h$ is almost surely given by $[0,4d] + \mbox{\rm supp}\,\rho$, in particular, $\min \sigma(h)=\omega_{\rm min}$, and $E=\omega_{\rm min}$ is almost surely not an eigenvalue of $h$.

We start by noting that Proposition~\ref{prop:FML} extends to infinite volume. By this we mean that the infinite volume Green function $G(x,y;z) := \langle \delta_x, (h-z)^{-1} \delta_y \rangle$ satisfies a fractional moment bound of the type (\ref{eq:FML}) for all $x,y \in \Z^d$ and uniformly in $\varepsilon \not= 0$ in the same regimes as described in Proposition~\ref{prop:FML}. This can be found in the same references as given for the finite volume case. Note that in infinite volume one can not allow $\varepsilon=0$ as real energies will almost surely be contained in the spectrum of $h$ (in situations other than trivial ones).

Another known fact is that Proposition~\ref{prop:EFCbounds}(a) extends to eigenfunction correlators for the infinite volume Anderson model.

\begin{proposition} \label{prop:EFCboundIV}
Assume that $h_L$ has localized $s$-moments in an interval $I$ for some $s\in (0,1)$. Then
\begin{equation} \label{eq:EFCboundIV}
\E \left( \sup_{{\tiny u:\R\to\C \;{\rm Borel}, \,|u|\le 1}} | \langle \delta_x, u(h) \chi_J(h) \delta_y \rangle| \right) \le C |J|^{1/(2-s)} e^{-\mu|x-y|/(2-s)}
\end{equation}
for every bounded open subinterval $J$ of $I$ and all $x, y \in \Z^d$. Here $C$ and $\mu$ are the constants from Proposition~\ref{prop:EFCbounds}(a).
\end{proposition}

That (\ref{eq:EFCboundIV}) follows by taking the $L\to\infty$ limit in (\ref{eq:EFCbound1}) is essentially a consequence of strong resolvent convergence, see \cite{ASFH, Stolz11} for details. The restriction to Borel functions $u$ and open intervals $J$ is due to measure theoretic requirements of the functional calculus and, in particular, the use of Lusin's theorem in the proof.

All the remaining results in this Appendix are most interesting for the case $\omega_{\rm min} =0$ (for $\omega_{\rm min}>0$ our results are obvious or easier to prove). This is the most interesting case for our applications, as it corresponds to gapless oscillator systems. It is also the most difficult case, as in this case operators such as $h^{-1/2}$, used to define singular eigenfunction correlators, are almost surely unbounded. We start by stating a lemma to clarify the meaning of $h^{-1/2}$ in this case.

For this we may consider any non-negative self-adjoint operator $h$ in a separable Hilbert space ${\mathcal H}$ such that $0$ is not an eigenvalue of $h$. Thus $h:D(h) \to R(h)$ is injective and $h^{-1}: R(h)\to D(h)$ is a well defined (but generally not bounded) linear operator. One may also express $h^{-1}$ in terms of the functional calculus of self-adjoint operators. For this let $E(\cdot)$ be the spectral resolution of $h$ and let $\varphi:\R\to \R$ be defined by
\begin{equation}
\varphi(t) = \left\{ \begin{array}{ll} 1/t, & \mbox{if $t>0$}, \\ 0, & \mbox{if $t \le 0$}. \end{array} \right.
\end{equation}
Then a self-adjoint operator $\varphi(h)$ in ${\mathcal H}$ is defined by the functional calculus (e.g.\ Chapter 7 of \cite{Weidmann}) as
\begin{equation}
D(\varphi(h)) = \{ f\in {\mathcal H}: \int_{(0,\infty)} \frac{1}{t^2}\, d\|E(t)f\|^2 < \infty \}, \quad \varphi(h)f = \int_{(0,\infty)} \frac{1}{t}\,dE(t)f.
\end{equation}

\begin{lemma} \label{lem:hinv}
(a) $h^{-1} = \varphi(h)$.

(b) $\varphi(h)$ is a non-negative self-adjoint operator and its unique non-negative square root $(\varphi(h))^{1/2}$ is given by $\psi(h)$, where
\begin{equation}
\psi(t) = \left\{ \begin{array}{ll} t^{-1/2}, & \mbox{if $ t>0$}, \\ 0, & \mbox{if $t \le 0$}. \end{array} \right.
\end{equation}

(c) The unique non-negative square root $h^{1/2}$ of $h$ is invertible and $\psi(h) = (h^{1/2})^{-1}$.
\end{lemma}

\begin{proof} We freely use background from the functional calculus of unbounded self-adjoint operators, see e.g.\ Section~7.2 of \cite{Weidmann} (note that the notation $\hat{E}(u)$ used there corresponds to $u(h)$ in the case where $E(\cdot)$ is the spectral family associated with $h$).

(a) The functional calculus allows to write $h =  {\rm id}(h)$ with the identity function ${\rm id}(t)=t$ for all $t\in \R$. We have $\varphi \cdot {\rm id} = \chi_{(0,\infty)}$ and thus $(\varphi \cdot {\rm id})(h) = \chi_{(0,\infty)}(h) = h$ as $h\ge 0$ and $0$ is not an eigenvalue of $h$. By Theorem~7.14(h) in \cite{Weidmann} we thus get  $\varphi(h) h \subset I$ and $D(\varphi(h)h) = D(h)$. The first fact implies $\varphi(h) \subset h^{-1}$, while the second gives $D(h^{-1}) = R(h) \subset D(\varphi(h))$. Combined this yields the claim.

(b) $\varphi\ge 0$ implies $\varphi(h)\ge 0$ by Theorem~7.14(f) of \cite{Weidmann}. It remains to show that $\psi(h)^2 = \varphi(h)$. To this end, we first observe that $D(\varphi(h)) \subset D(\psi(h))$. This is the same as saying that $\int_{(0,\infty)} t^{-2}\,d\|E(t)f\|^2<\infty$ implies $\int_{(0,\infty)} t^{-1}\,d\|E(t)f\|^2<\infty$, which follows from Cauchy-Schwarz.
Now, as $\psi^2 = \varphi$, it follows from Theorem~7.14(h) in \cite{Weidmann} that $\psi(h)^2 \subset \varphi(h)$ with $D(\psi(h)^2) = D(\psi(h)) \cap D(\varphi(h)) = D(\varphi(h))$, giving the claim.

(c) This follows from an application of Theorem~7.14(f) of \cite{Weidmann} similar to the one in part (a), this time to the product of the functions $\psi$ and $g$, where $g(t)= t^{1/2}$ for $t\ge 0$ and $g(t)=0$ for $t<0$.

\end{proof}

Based on these facts we will from now on write $h^{-1/2}$ for $\psi(h)$, which means that we have defined $h^{-1/2} = (h^{-1})^{1/2}$. This notation is further justified by Lemma~\ref{lem:hinv}(c), which says that $h^{-1/2} = (h^{1/2})^{-1}$.

We now return to studying the infinite volume Anderson model $h$.

\begin{lemma} \label{lem:deltaindomain}
Let $h$ be the Anderson model on $\Z^d$ with assumptions as above. Also let $x\in \Z^d$. Then $\delta_x \in D(h^{-1/2})$ almost surely.
\end{lemma}

\begin{proof}
By Lemma~\ref{lem:hinv}(b) we have to show that $\int_{(0,\infty)} t^{-1}\,d\|E(t) \delta_x\|^2 <\infty$ almost surely. This will follow from
\begin{equation} \label{eq:tobeshown}
\E \left( \left( \int_{(0,\infty)} t^{-1}\,d\|E(t) \delta_x\|^2 \right)^{1/2} \right) < \infty.
\end{equation}

By the infinite volume version of Proposition~\ref{prop:FML}(a) for $s=1/2$ we have
\begin{equation} \label{eq:fact1}
\E \left( |\langle \delta_x, (h-i\varepsilon)^{-1} \delta_x \rangle|^{1/2} \right) \le C < \infty
\end{equation}
uniformly in $\varepsilon>0$.

Using that $0$ almost surely is not an eigenvalue of $h$, we get that, almost surely,
\begin{eqnarray} \label{eq:fact2}
{\rm Re} \langle \delta_x, (h-i\varepsilon)^{-1} \delta_x \rangle & = & {\rm Re} \int_{(0,\infty)} \frac{1}{t-i\varepsilon} d\|E(t) \delta_x\|^2 \\
& = & \int_{(0,\infty)} \frac{t}{t^2+\varepsilon^2} d\|E(t) \delta_x\|^2. \nonumber
\end{eqnarray}
For all $t>0$ we have $t/(t^2+\varepsilon^2) \nearrow 1/t$ as $\varepsilon \searrow 0$. Thus the monotone convergence theorem shows that
\begin{equation}
\int_{(0,\infty)} \frac{t}{t^2+\varepsilon^2} d\|E(t) \delta_x\|^2 \nearrow \int_{(0,\infty)} \frac{1}{t} d\|E(t) \delta_x\|^2,
\end{equation}
as well as
\begin{equation} \label{eq:mctresult}
\E \left( \left( \int_{(0,\infty)} \frac{1}{t} d\|E(t) \delta_x\|^2 \right)^{1/2} \right) = \lim_{\varepsilon \searrow 0} \E \left( \left( \int_{(0,\infty)} \frac{t}{t^2+\varepsilon^2} d\|E(t) \delta_x\|^2 \right)^{1/2} \right).
\end{equation}
Combining (\ref{eq:fact1}) and (\ref{eq:fact2}) (and $|{\rm Re}\,z| \le |z|$) we see that the right hand side of (\ref{eq:mctresult}) is uniformly bounded in $\varepsilon>0$. This yields (\ref{eq:tobeshown}).

\end{proof}

We can now state and prove infinite volume versions of parts (b) and (c) of Proposition~\ref{prop:EFCbounds}. While more general results hold, we only do this for $\alpha=-1/2$, the relevant case for our applications.

\begin{proposition} \label{prop:EFCboundIVbc}
(a) If $h_L$ is $s$-localized in $[0,E_0]$ for some $s\in (2/3,1)$, then there exists $C_1<\infty$ such that
\begin{equation} \label{eq:EFCboundIV2}
\E \left( \sup_{{\tiny u \;{\rm Borel}, \,|u|\le 1}} |\langle \delta_x, h^{-1/2} u(h) \chi_{[0,E_0]}(h) \delta_y \rangle |\right) \le C_1 e^{-\mu|x-y|/(2-s)}
\end{equation}
for all $x,y \in \Z^d$. Here $\mu>0$ is as in Proposition~\ref{prop:EFCboundIV}.

(b) Suppose that $\varphi:(0,\infty) \to \C$ satisfies $\varphi(t) \le Ct^{-1/2}$ for $t$ near $0$, and that $\varphi$ has an analytic extension to a semi strip $\{z: {\rm Re}\,z>0, \, |{\rm Im}\,z| < \eta\}$ for some $\eta>0$. Then there exist $C_2< \infty$ and $\mu_2>0$ such that
\begin{equation} \label{prop:EFCboundIV3}
\E \left( | \langle \delta_x, \varphi(h) \delta_y \rangle| \right) \le C_2 e^{-\mu_2|x-y|}
\end{equation}
for all $x,y \in \Z^d$.
\end{proposition}

\begin{proof}
As $0$ is almost surely not an eigenvalue of $h$ and $h^{-1/2}$ commutes with the bounded operator $u(h) \chi_{[0,E_0]}(h)$, the vector $h^{-1/2} u(h)\chi_{[0,E_0]}(h) \delta_y$ is almost surely well-defined by Lemma~\ref{lem:deltaindomain}. Thus the left hand side of (\ref{eq:EFCboundIV2}) makes sense. The same can be said for the left hand side of (\ref{prop:EFCboundIV3}) as we can write $\varphi(h) = h^{-1/2} h^{1/2} \varphi(h)$ and, by assumption, $h^{1/2} \varphi(h)$ is bounded.

The proof of (a) now proceeds by essentially the same Riemann sum argument as the proof of Proposition~\ref{prop:EFCbounds}(b). We use Proposition~\ref{prop:EFCboundIV} instead of Proposition~\ref{prop:EFCbounds}(a), which requires to decompose $[0,E_0]$ into open intervals $I_n := (E_0/(n+1), E_0/n)$. This does not impact the proof as the countable set $\{0\} \cup \{E_0/n: n=1,2,\ldots\}$ almost surely carries no spectral measure of $h$. The only other part of the proof which needs a bit more care than before is (\ref{eq:crucialsplit}). Here we use again that almost surely $\delta_y \in D(h^{-1/2})$. Thus, almost surely,
\begin{equation}
h^{-1/2} u(h) \chi_{[0,E_0]}(h) \delta_y = \sum_n h^{-1/2} u(h) \chi_{I_n}(h) \delta_y
\end{equation}
in the sense of strong convergence. The rest of the argument is unchanged.

To prove (b), we first note that we may assume that $|\varphi(t)| \le C t^{-1/2}$ not just near $0$, but uniformly on all the spectra of $h$ and $h_L$, which is due to the uniform boundedness of these spectra.

Assume that $0$ is not an eigenvalue of $h$ and that $\delta_y \in D(h^{-1/2})$, which, by Lemma~\ref{lem:deltaindomain} holds almost surely. By Lemma~\ref{lem:hinv} this means $\int_{(0,\infty)} t^{-1}\,d\|E(t) \delta_y\|^2 <\infty$. It follows that $\int_{(0,\infty)} |\varphi(t)|^2\,d\|E(t) \delta_y\|^2 < \infty$, i.e.\ $\delta_y \in D(\varphi(h))$.

For $\delta>0$ define continuous functions $\varphi_{\delta}:\R \to \C$ such that $\varphi_{\delta}(t) = \varphi(t+\delta)$ for $t\ge 0$ (this is possible as $\delta>0$ allows to avoid the singularity of $\varphi$ at $0$, while the exact choice of $\varphi_{\delta}$ on $(-\infty,0)$ is irrelevant).

We claim that
\begin{equation} \label{eq:nlim}
\varphi(h) \delta_y = \lim_{n\to\infty} \varphi_{1/n}(h) \delta_y.
\end{equation}
To see this, note that by the functional calculus
\begin{equation}
\|(\varphi(h) - \varphi_{1/n}(h))\delta_y\|^2 = \int_{(0,\infty)} |\varphi(t) - \varphi(t+1/n)|^2 \,d\|E(t) \delta_y\|^2.
\end{equation}
We have $\varphi(t+1/n) \to \varphi(t)$ for all $t>0$ by continuity as well as
\begin{equation}
|\varphi(t)-\varphi(t+1/n)|^2 \le 2(|\varphi(t)|^2 + |\varphi(t+1/n)|^2) \le \frac{4C^2}{t}
\end{equation}
on $\sigma(h)$. The latter is integrable with respect to $d\|E(t) \delta_y\|^2$. Thus (\ref{eq:nlim}) follows by dominated convergence.

Also, for fixed $n$ the function $\varphi_{1/n}$ is continuous and thus it follows from strong resolvent convergence that
\begin{equation} \label{eq:Llim}
\varphi_{1/n}(h) \delta_y = \lim_{L\to\infty} \varphi_{1/n}(h_L) \delta_y.
\end{equation}

It remains to show the existence of $C_2<\infty$ and $\mu_2>0$ such that
\begin{equation} \label{eq:Lnuniformbound}
\E \left( |\langle \delta_x, \varphi_{1/n}(h_L) \delta_y \rangle| \right) \le C_2 e^{-\mu_2 |x-y|}
\end{equation}
for all $x, y \in \Z^d$, uniformly in $L$ and $n$. For (\ref{eq:nlim}) and (\ref{eq:Llim}) combine to give
\begin{equation}
|\langle \delta_x, \varphi(h) \delta_y \rangle| = \lim_{n\to\infty} \lim_{L\to\infty} |\langle \delta_x, \varphi_{1/n}(h_L) \delta_y \rangle|
\end{equation}
almost surely. Thus (\ref{prop:EFCboundIV3}) follows from (\ref{eq:Lnuniformbound}) and Fatou's lemma.

The proof of (\ref{eq:Lnuniformbound}) is done by a straightforward inspection of the proof of Proposition~\ref{prop:EFCbounds}(c), showing that one may replace $\varphi$ by $\varphi_{1/n}$ and gets bounds uniform in $n$. This comes down to two instants in the proof. In the part corresponding to the spectral projection on $[0,E_0]$ use that
\begin{equation}
\varphi_{1/n}(h_L) \chi_{[0,E_0]}(h_L) = h_L^{-1/2} \psi_n(h_L) \chi_{[0,E_0]}(h_L),
\end{equation}
where $|\psi_n(t)| = |t^{1/2} \varphi_{1,n}(t)| \le C$. Thus one can use Proposition~\ref{prop:EFCbounds}(b) uniformly in $n$. Second, when bounding $\E(|\langle \delta_x, \varphi_{1/n}(h_L) \chi_{[E_0,\infty)}(h_L) \delta_y \rangle|)$ all one needs to observe is that the constants
\begin{equation}
C_n = \frac{1}{2\pi} \max \,\{ |\varphi_{1/n}(z)|: z\in \Gamma \}
\end{equation}
entering (\ref{eq:Greenbound}) are uniformly bounded in $n$.

\end{proof}

It is now straightforward to adapt Proposition~\ref{prop:verysingEFC} and its proof to infinite volume. All that needs to be done to ensure that the infinite volume quantities are well-defined is to distribute singular operator functions evenly to both sides of the inner products.

\begin{proposition} \label{prop:verysingEFCIV}

(a) If $h_L$ has $s$-localized moments in $[0,E_0]$ for some $s\in (2/3,1)$, then there exist $C_3<\infty$ and $\mu_3>0$ such that
\begin{equation} \label{eq:SEFCbound1IV}
\E \left( \sup_{{\tiny u \;{\rm Borel}, \,|u|\le 1}} | \langle h^{-1/2} \delta_x, h^{-1/2} u(h) \chi_{[0,E_0]}(h) \delta_y \rangle |^{1/2} \right) \le C_3 e^{-\mu_3 |x-y|}
\end{equation}
for all $x,y \in \Z^d$.

(b) Suppose that $\varphi_1$ and $\varphi_2$ have the properties of $\varphi$ in Proposition~\ref{prop:EFCboundIVbc}(b). Then there exist $C_4<\infty$ and $\mu_4>0$ such that
\begin{equation} \label{eq:EFCbound4IV}
\E \left( |\langle \varphi_1(h) \delta_x, \varphi_2(h) \delta_y \rangle |^{1/2} \right) \le C_4 e^{-\mu_4 |x-y|}
\end{equation}
for all $x,y \in \Z^d$.

\end{proposition}

\end{document}